\documentclass[final,3p,times,square]{elsarticle}

\usepackage{amssymb}
\usepackage{amsmath}
\usepackage{graphicx}
\usepackage{dcolumn}
\usepackage{bm}
\usepackage[utf8]{inputenc}
\usepackage[T1]{fontenc}
\usepackage{etoolbox}
\usepackage{booktabs} 
\usepackage{array} 
\usepackage{multirow} 
\usepackage{makecell} 
\usepackage{threeparttable}
\usepackage{tabularx}
\usepackage{float} 
\usepackage{url}
\usepackage{algorithm}
\usepackage{algorithmic}
\usepackage{amsthm}
\usepackage{colortbl}
\usepackage {color}
\usepackage{xcolor}
\usepackage{appendix}
\usepackage{lineno}

\journal{Journal of Computational Physics}

\begin{document}
\begin{frontmatter}

\title{An explainable operator approximation framework under the guideline of Green's function}

\author[label1,label2]{Jianghang Gu}
\author[label1]{Ling Wen}
\author[label2,label3]{Yuntian Chen*} \cortext[cor1]{ychen@eitech.edu.cn}
\author[label1,label2,label3]{Shiyi Chen}

\affiliation[label1]{organization={State Key Laboratory of Turbulence and Complex Systems, College of Engineering, Peking University},
            city={Beijing},
            postcode={100871},
            state={},
            country={P. R. China}}

\affiliation[label2]{organization={Ningbo Institute of Digital Twin, Eastern Institute of Technology},
            city={Ningbo},
            postcode={315200},
            state={Zhejiang},
            country={P. R. China}}
            
\affiliation[label3]{organization={Zhejiang Key Laboratory of Industrial Intelligence and Digital Twin, Eastern Institute of Technology},
            city={Ningbo},
            postcode={315200},
            state={Zhejiang},
            country={P. R. China}}

\begin{abstract}
\textcolor{black}{Compared to traditional methods such as the finite element and finite volume methods, the Green's function approach offers the advantage of providing analytical solutions to linear partial differential equations (PDEs) with varying boundary conditions and source terms, without the need for repeated iterative solutions. Nevertheless, deriving Green's functions analytically remains a non-trivial task.}
In this study, we \textcolor{black}{develop a framework inspired by the architecture of deep operator networks (DeepONet) to learn embedded Green's functions and solve PDEs through integral formulation, termed the Green's operator network (GON)}. 
Specifically, the Trunk Net within GON is designed to approximate the unknown Green's functions of the system, while the Branch Net are utilized to approximate the auxiliary gradients of the Green's function. These outputs are subsequently employed to perform surface integrals and volume integrals, incorporating user-defined boundary conditions and source terms, respectively.
The effectiveness of the proposed framework is demonstrated on three types of PDEs in 3D bounded domains: Poisson equations, reaction-diffusion equations, and Stokes equations. Comparative results in these cases demonstrate that GON's accuracy and generalization ability surpass those of existing methods, including Physics-Informed Neural Networks (PINN), DeepONet, Physics-Informed DeepONet (PI-DeepONet), and Fourier Neural Operators (FNO).
Code and data is available at \url{https://github.com/hangjianggu/GreensONet}.
\end{abstract}

\end{frontmatter}

\section{Introduction}
\label{sec1}

Traditional computational methods, such as Finite Difference Methods (FDM) \cite{perrone1975general}, Finite Element Methods (FEM) \cite{brenner2004finite} and Finite Volume Methods (FVM) \cite{eymard2000finite}, are well-established for solving partial differential equations (PDEs) and have demonstrated robustness in various engineering applications. However, when applied to tasks requiring repeated forward problem evaluations under varying initial or boundary conditions, such as parameter optimization and inverse problem solving, these methods can be computationally expensive \cite{lu2021learning, lifourier}. 
Recent advances in deep learning have shown potential in certain problem settings \cite{pinkus1999approximation, cai2021physics, wang2021learning}. While these approaches are not yet a replacement for classical numerical methods, they offer an intriguing direction for exploring alternative solution strategies \cite{aldirany2024multi, chen2021theory, zhang2024filtered, mouratidou2024ensemble, li2024probabilistic, garg2019enhanced}.

Prominent deep learning methods for solving PDEs can be broadly categorized into three approaches: (1) learning the solution to the PDEs directly \cite{zang2020weak, raissi2019physics, yu2018deep}, (2) learning the operators that define the underlying physics \cite{lu2021learning, li2020fourier, li2020neural}, and (3) learning the operators with PDE-based loss constraints \cite{wang2021learning,liu2023deepoheat, lu2024fast}. Each approach offers distinct advantages and limitations.
The first approach focuses on approximating the solution to the PDEs by directly minimizing the residual of the governing equations. A notable example is physics-informed neural networks (PINNs) \cite{raissi2019physics}, which embed the PDEs into the loss function of the neural network. 
This method effectively enforces physical laws without requiring labeled data but can struggle with convergence in complex or high-dimensional problems.
The second approach shifts focus to learning the differential operators that describe the underlying physics. Methods such as Fourier Neural Operators (FNO) \cite{li2020fourier} and Deep Operator Networks (DeepONet) \cite{lu2021learning} aim to approximate the operator mapping input parameters to solutions for a family of PDEs. These methods enable efficient predictions for new parameters within the training range, but their accuracy may degrade for out-of-distribution parameters or complex domains.
The third approach combines elements of the first two, integrating operator learning with physics-based constraints. For example, physics-informed DeepONet (PI-DeepONet) \cite{wang2021learning} leverages the strengths of PINN and DeepONet by simultaneously learning the operator and minimizing the PDEs residual. This hybrid approach improves generalization over parameter families while retaining physical consistency. 
However, it often requires significant computational resources due to the combined loss constraints.

The primary objective of this paper is to develop an operator approximation framework with physical interpretability. Central to the framework is the direct approximation of the Green’s function for PDE systems, enabling solutions to be obtained via integration. This approach guarantees generalization performance, and its utility in analyzing the well-posedness and regularity properties of PDEs.
A series of prior works have explored the use of Green’s function approximations for solving PDEs \cite{li2020neural, gin2021deepgreen, boulle2022data, teng2022learning, aldirany2024operator, negi2024learning}. We extend these efforts by developing a more general framework that is boundary-invariant and source-term-invariant, capable of handling 3D bounded domains.
Below, we provide a brief overview of these methods.
The authors in \cite{li2020neural} introduced the graph neural operator, which is inspired by the Green’s function. 
In \cite{gin2021deepgreen}, a dual-autoencoder architecture is presented to approximate the operator for non-linear boundary value problems, by linearizing the problem and approximating the corresponding Green’s function. Nevertheless, the linear integral operator is also given by the neural network, which introduces redundancy and compromises accuracy.
Boulle et al. \cite{boulle2022data} tackled PDE problems by decomposing them into two components: one with homogeneous boundary conditions and the other with non-homogeneous boundary conditions. They employed two separate rational neural networks \cite{boulle2020rational} to approximate the Green's function influenced by force terms and the homogeneous solution influenced by boundary conditions, respectively. However, this approach requires retraining the networks whenever boundary conditions change, limiting its flexibility.
Teng et al. \cite{teng2022learning} proposed an unsupervised method, akin to PINN, to learn the Green’s functions for linear reaction-diffusion equations with Dirichlet boundary conditions. Their approach approximates the Dirac delta function with a Gaussian density function. 
Building on this, Negi et al. \cite{negi2024learning} used a radial basis function (RBF) kernel-based neural network to better adapt to the singularity of the Dirac delta function. 
While these methods improve the approximation of Green’s functions, the use of surrogate functions like Gaussian or RBF kernels imposes limitations on accuracy, ultimately affecting the precision of the PDE solutions.
Another line of physics-informed frameworks for learning Green’s functions was proposed by Aldirany et al. \cite{aldirany2024operator}, who introduced a physics-informed DeepONet trained by minimizing the residuals of the governing PDEs, along with initial and boundary conditions across a family of problems.
Their formulation offers a novel integration of physics-constraints into neural operator learning, advancing the capability to solve families of PDEs with limited supervision.
One limitation of this approach is its reliance on fixed boundary conditions, which may affect its flexibility in handling more diverse problem classes.
In conclusion, these methods demonstrate significant progress in using neural networks to approximate Green’s functions and provide efficient solutions for PDEs. However, they lack a boundary-invariant and source-term-invariant framework to handle  3D bounded domains. \textcolor{black}{The proposed method aims to address this gap, with the present study focusing on Dirichlet boundary conditions as representative cases.}

In this paper, we propose a general Green's function approximation framework based on the structure of DeepONet, denoted as \textbf{GON}. DeepONet serves as the backbone for our framework due to its versatility, as it does not rely on prior knowledge of the solution structure and can be readily applied to a wide range of problems. 
In GON, the Trunk Net is designed to approximate the unknown Green's functions of the system, while the Branch Net are employed to estimate the auxiliary gradients of the Green’s function. These outputs are then used to perform surface and volume integrals, incorporating user-defined boundary conditions and source terms. By minimizing the deviation between the solutions derived from the acquired Green's functions and the exact solutions, the framework effectively tunes the Green's function for accurate approximation.
To further enhance the capability of the Trunk and Branch Net in capturing the singularities of Green’s functions, we introduce a novel binary-structured neural network architecture within these components.
This design improves the accuracy of the approximations. 
GON supports flexible geometric inputs, such as meshes or scattered data points, and accommodates versatile boundary conditions, source terms, and different types of equations, including heat conduction equations and reaction-diffusion equations. 
Furthermore, it extends to \textcolor{black}{vectorial problem}, such as Stokes equations. 
Our experiments demonstrate that GON consistently outperforms state-of-the-art methods, including \textcolor{black}{previous Green's function based networks \cite{boulle2022data, teng2022learning, aldirany2024operator}}, PINN \cite{raissi2019physics}, DeepONet \cite{lu2021learning}, PI-DeepONet \cite{wang2021learning}, and FNO \cite{li2020fourier}, across several classical 3D PDE benchmark cases, showcasing its superior performance and broad applicability.

Our contributions are summarized as follows:
\begin{itemize}
\item We develop a general Green's function-based deep learning framework capable of learning boundary-invariant and source-term invariant Green's functions in 3D bounded domains.
\item We introduce a computing approach for efficiently calculating the convolution between the learned Green’s function and the loading function, supporting versatile geometric domains, boundary conditions, homogeneous and heterogeneous PDE coefficients.  
\item We employ a binary-structured neural network to effectively capture the singularities of Green’s functions, ensuring improvement in approximation accuracy.
\end{itemize}

This paper is organized into four sections. In Section \ref{sec2}, we present the fundamental mathematical theorems as preliminaries. Section \ref{sec3} introduces the GON framework. In Section \ref{sec4}, we present numerical results and validations. Finally, conclusions and outlooks are provided in Section \ref{sec5}.

\section{Preliminaries\label{sec2}}
We introduce here some preliminaries and notations in order to describe the notion of the operators in PDEs. 
We first present the model problem and continue with a brief account of Green's functions to solve boundary-value problems.
Based on these foundations, we outline the core approach of our framework.

Let $\Omega \subset \mathbb{R}^d$ be a bounded domain, we consider the linear PDE operator with Dirichlet boundary condition of the following form:
\begin{equation}
    \left\{\begin{aligned}
    \mathcal{L}(u)(\boldsymbol{x})=f(\boldsymbol{x}), \quad &\boldsymbol{x} \in \Omega \\
    u(\boldsymbol{x})=g(\boldsymbol{x}), \quad &\boldsymbol{x} \in \partial \Omega 
    \end{aligned}\right.
    \label{eq1}
\end{equation}
where $f(\boldsymbol{x})$ is the given source term, $g(\boldsymbol{x})$ is the boundary value. \textcolor{black}{The linear differential operator \(\mathcal{L}\) is defined as  
$
\mathcal{L}(u)(\boldsymbol{x}) = -\nabla \cdot \left(a(\boldsymbol{x}) \nabla u(\boldsymbol{x})\right) + r(\boldsymbol{x}) u(\boldsymbol{x}),
$
where \(a(\boldsymbol{x})\) and \(r(\boldsymbol{x})\) are material coefficients.}

The Green's function $G(\boldsymbol{x}, \boldsymbol{\xi})$ represents the impulse response of the PDE subject to homogeneous Dirichlet boundary condition, that is, for any impulse source point $\boldsymbol{\xi} \in \Omega$,
\begin{equation}
    \left\{\begin{aligned}
    \mathcal{L}(G)(\boldsymbol{x}, \boldsymbol{\xi})=\delta(\boldsymbol{x}-\boldsymbol{\xi}), \quad & \boldsymbol{x} \in \Omega \\
    G(\boldsymbol{x}, \boldsymbol{\xi})=0, \quad & \boldsymbol{x} \in \partial \Omega
    \end{aligned}\right.
    \label{eq2}
\end{equation}
where $\delta(\boldsymbol{x})$ denotes the Dirac delta source function satisfying $\delta(\boldsymbol{x})=0$ if $\boldsymbol{x} \neq 0$ and $\int_{\mathbb{R}^d} \delta(\boldsymbol{x}) d \boldsymbol{x}=$ 1. \textcolor{black}{Note that the Green’s function formulation in Eq. \eqref{eq2} holds when the differential operator \( \mathcal{L} \) in Eq. \eqref{eq1} is self-adjoint. If the bilinear form associated with \( \mathcal{L} \) lacks symmetry, such as in the presence of a convective term, the corresponding operator governing the Green’s function would differ from Eq. \eqref{eq2}.}  

The Green's function \( G(\boldsymbol{x}, \boldsymbol{\xi}) \) satisfies Eq. (\ref{eq2}) independently of the specific boundary conditions and force term in Eq. (\ref{eq1}). 
If \( G(\boldsymbol{x}, \boldsymbol{\xi}) \) in Eq. (\ref{eq2}) is known, the solution to the problem defined by Eq. (\ref{eq1}) can be directly computed using the following formula, which accommodates both variable boundary conditions, source terms and \textcolor{black}{material parameters $a(\boldsymbol{\xi})$}:  
\begin{equation}
    u(\boldsymbol{x})=\int_{\Omega} f(\boldsymbol{\xi}) G(\boldsymbol{x}, \boldsymbol{\xi}) d \boldsymbol{\xi} - \int_{\partial \Omega} g(\boldsymbol{\xi}) a(\boldsymbol{\xi})\left(\nabla_{\boldsymbol{\xi}} G(\boldsymbol{x}, \boldsymbol{\xi}) \cdot \mathbf{n}_{\boldsymbol{\xi}}\right) d S(\boldsymbol{\xi}). \quad \forall \boldsymbol{x} \in \Omega
    \label{eq3}
\end{equation}
\textcolor{black}{While the framework is applicable to a broad class of boundary conditions—including Neumann and mixed types—this study focuses on the Dirichlet case for clarity and ease of validation. Extension to other boundary conditions is theoretically feasible by incorporating additional boundary integrals into the representation in Eq.(\ref{eq3}).
}

Green’s functions can be obtained analytically via eigenfunction expansions or numerically  solving a singular PDE (e.g., by approximating the Dirac delta function). 
However, when the geometry of the domain is complex, or the PDE has variable coefficients, finding the analytical form of the Green’s function is a non-trivial task \cite{teng2022learning, aldirany2024operator, boulle2020rational}.
In this study, we attempt to utilize our GON framework to approximate the unknown Green's function numerically.

The method for discovering Green's functions of scalar differential operators can be extended naturally to systems of differential equations. Let $f=\left[\begin{array}{lll}f^1 & \cdots & f^{N_f}\end{array}\right]^{\top}: \Omega \rightarrow \mathbb{R}^{N_f}$ be a vector of $N_f$ forcing terms, \textcolor{black}{, $g=\left[\begin{array}{lll}g^1 & \cdots & g^{N_u}\end{array}\right]^{\top}: \Omega \rightarrow \mathbb{R}^{N_u}$ be a vector of $N_u$ boundary conditions,} and $u=\left[\begin{array}{lll}u^1 & \cdots & u^{N_u}\end{array}\right]^{\top}: \Omega \rightarrow \mathbb{R}^{N_u}$ be a vector of $N_u$ system responses such that

\begin{equation}
    \mathcal{L}\left[\begin{array}{c}
    u^1 \\ \vdots \\ u^{N_u}
    \end{array}\right]=
    \left[\begin{array}{c}
    f^1 \\ \vdots \\ f^{N_f}
    \end{array}\right], 
    \quad \mathcal{D}\left(
    \left[\begin{array}{c}
    u^1 \\ \vdots \\ u^{N_u}
    \end{array}\right], 
    \Omega\right)=\left[\begin{array}{c}
    g^1 \\ \vdots \\ g^{N_u}
    \end{array}\right],
\end{equation}
where $\mathcal{D}$ is a linear operator acting on the boundary.

We can express the relation between the system’s response and the forcing term using Green’s functions as an integral formulation, i.e.,
\begin{equation}
    u^i(\boldsymbol{x})=\sum_{j=1}^{N_f} \int_{\Omega} G_{i, j}(\boldsymbol{x}, \boldsymbol{\xi}) f^j(\boldsymbol{\xi}) \mathrm{d} \boldsymbol{\xi} -\int_{\partial \Omega} g^i(\boldsymbol{\xi}) a^i(\boldsymbol{\xi})\left(\nabla_{\boldsymbol{\xi}} G_{i, j}(\boldsymbol{x}, \boldsymbol{\xi}) \cdot \mathbf{n}_{\boldsymbol{\xi}}\right) d S(\boldsymbol{\xi}), \quad x \in \Omega,
\end{equation}
for $1 \leq i \leq N_u$. \textcolor{black}{$a^i(\boldsymbol{\xi})$ is a vector of material parameters.} $G_{i, j}: \Omega \times \Omega \rightarrow \mathbb{R} \cup\{ \pm \infty\}$ is a component of the Green's matrix for $1 \leq i \leq N_u$ and $1 \leq j \leq N_f$. Specifically, the $N_u \times N_f$ matrix of Green's functions can be written as:

\begin{equation}
    G(\boldsymbol{x}, \boldsymbol{\xi})=
    \left[\begin{array}{ccc}
    G_{1,1}(\boldsymbol{x}, \boldsymbol{\xi}) & \cdots & G_{1, N_f}(\boldsymbol{x}, \boldsymbol{\xi}) \\
    \vdots & \ddots & \vdots \\
    G_{N_u, 1}(\boldsymbol{x}, \boldsymbol{\xi}) & \cdots & G_{N_u, N_f}(\boldsymbol{x}, \boldsymbol{\xi})
    \end{array}\right]. \quad \boldsymbol{x}, \boldsymbol{\xi} \in \Omega.
    \label{eq6}
\end{equation}

Following Eq. (\ref{eq6}), we remark that the differential equations decouple, and therefore we can learn each row of the Green's function matrix independently. That is, for each row $1 \leq i \leq N_u$, we train $N_f$ neural networks to approximate the components $G_{i, 1}, \ldots, G_{i, N_f}$.

Our objective is to construct a neural network to approximate the Green's functions for linear PDE systems. This approach enables the efficient computation of solutions for varying source terms \( f(\boldsymbol{x}) \) and boundary conditions \( g(\boldsymbol{x}) \) by leveraging the approximated Green's function. 
To achieve this, the network is trained on a family of source terms \( f(\boldsymbol{x}) \) and boundary conditions \( g(\boldsymbol{x}) \) generated using Gaussian random fields (GRF) \cite{seeger2004gaussian}. Once trained, the resulting neural network provides a flexible and efficient framework for approximating solutions to the specified PDEs, accommodating previously unseen boundary conditions and source terms while preserving computational efficiency and generalization capability.

\section{Methodologies\label{sec3}}
\subsection{Framework of GON}
The DeepONet architecture is a versatile framework capable of addressing a broad range of problems with varying input parameters. 
Based on the structure of DeepONet, our approach aims at extracting the underlying Green's functions to capture the influence of loading terms on the solution. This formulation enables a more physically interpretable solution representation. Besides, it should be \textcolor{black}{noted} that the Green’s function is inherently dependent on the material parameters of the system (e.g., \( a(\boldsymbol{\xi}) \)), meaning that changes in these parameters necessitate a corresponding update in the Green’s function.
We demonstrate the effectiveness of our approach by solving three classical types of PDEs: \textcolor{black}{Poisson} equations, reaction-diffusion equations, and Stokes equations, for both homogeneous and heterogeneous coefficients in 3D bounded domains, as detailed in Section \ref{sec4}.

\begin{figure}[H]
    \centering
    \includegraphics[width=0.9\linewidth]{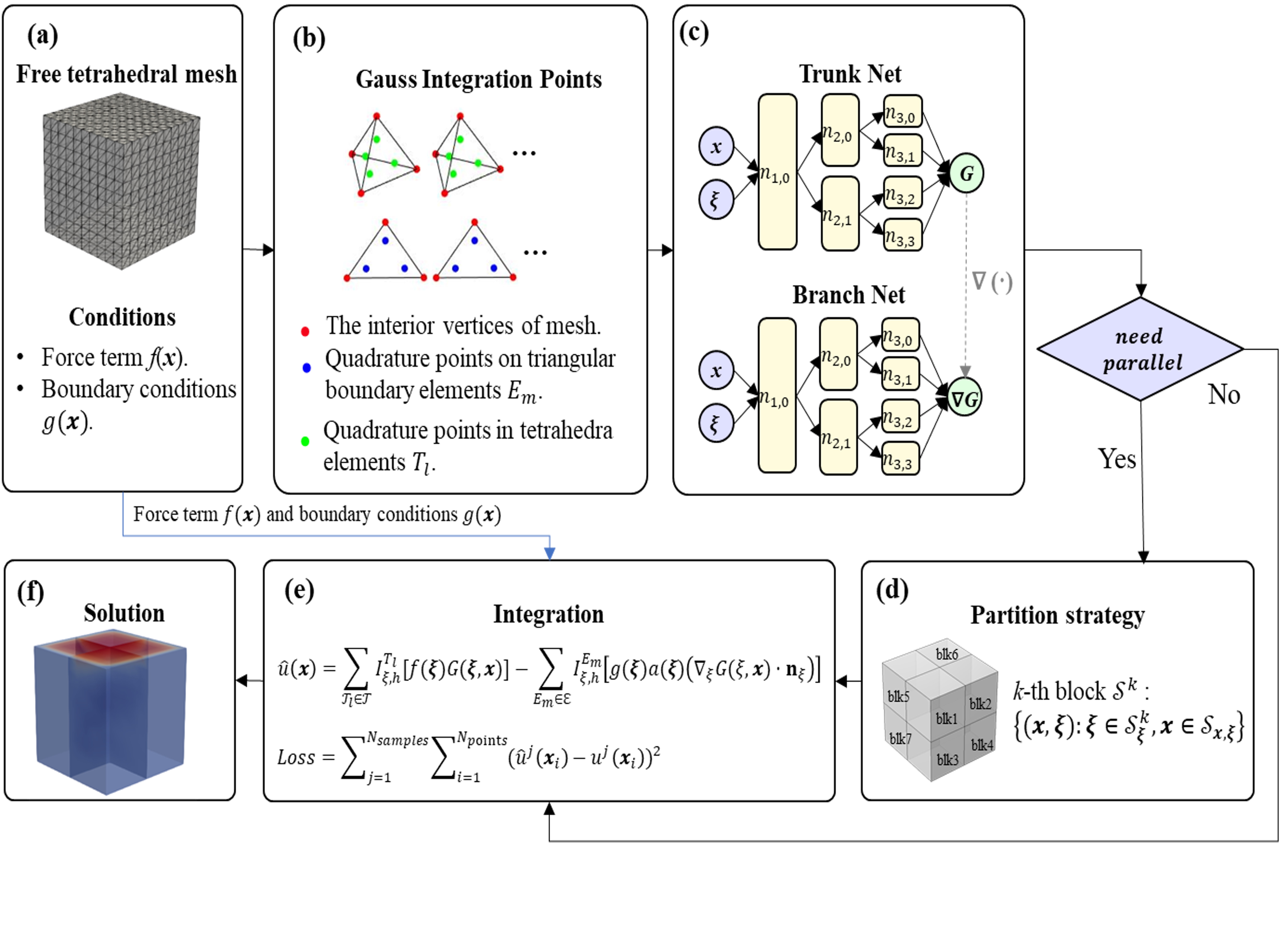}
    \caption{The framework of GON: (a) Import user-defined free tetrahedral mesh and user-defined physical conditions; (b) Calculate the locations of Gauss integration points and integration weights; (c) Constructions of the Trunk Net and Branch Net of the GON based on binary structured neural networks; (d) Domain partition and parallel computation strategy; (e) Volterra integration based on acquired Green's function; (f) The calculated solutions.}
    \label{fig:workflow}
\end{figure}

As shown in Figure \ref{fig:workflow}, the GON framework is designed to efficiently solve operator approximation problems by leveraging the principles of Green's functions, Volterra integral equations and DeepONet. 

The framework begins with the import of a user-defined mesh and physical conditions (step a), followed by the calculation of Gauss integration points and their corresponding weights (step b), which are crucial for accurate numerical integration. In this study, a 4-point Gauss integration rule is applied to all tetrahedral elements within the mesh, while a 3-point Gauss integration rule is used for the triangular boundary faces. Other Gauss integration rules are also available in our code. \textcolor{black}{It is indicate that} increasing the number of integration points can improve the accuracy. Our chosen integration scheme strikes a balance between computational efficiency and accuracy. \textcolor{black}{This trade-off strategy can be supported by the analysis presented in Appendix A.1.}

In step (c), inspired by DeepONet, the Trunk Net and Branch Net are constructed to facilitate the representation of the operator through hierarchical learning.
Furthermore, novel binary-structured neural networks are employed in both the Trunk Net and the Branch Net owing to their strong convergence properties\textcolor{black}{, which have been demonstrated in \cite{liu2024binary}}. 

To ensure computational efficiency, step (d) employs domain partitioning based on the principle of independent computation for the Green's function at each point. This approach is further enhanced by parallel computation strategies, enabling the efficient resolution of large-scale problems. 

In step (e), Volterra integration is performed using the Green's function obtained in the previous steps, enabling the calculation of the operator’s response. By minimizing the deviation between the computed and exact solutions, the Trunk Net and Branch Net are progressively trained until the deviation meets the desired tolerance. Instead of using iterative loops, we directly compute the numerical integration through matrix operations. 
For cases with up to 10,000 grid elements, we can compute the global integral in just 0.5 seconds.

Finally, in step (f), the framework outputs the calculated solutions. 

The structured design of GON leverages the physical principles of Green's function and the Volterra integral, enabling explicit modeling of system responses to localized perturbations. This design provides a physically interpretable representation of the learned operator. Powered by PaddlePaddle \cite{ma2019paddlepaddle}, GON efficiently approximates solutions to complex PDEs under varying conditions. The following subsections will present the technical details in depth.

\subsection{Training datasets}
\begin{figure}[H]
    \centering
    \includegraphics[width=0.7\linewidth]{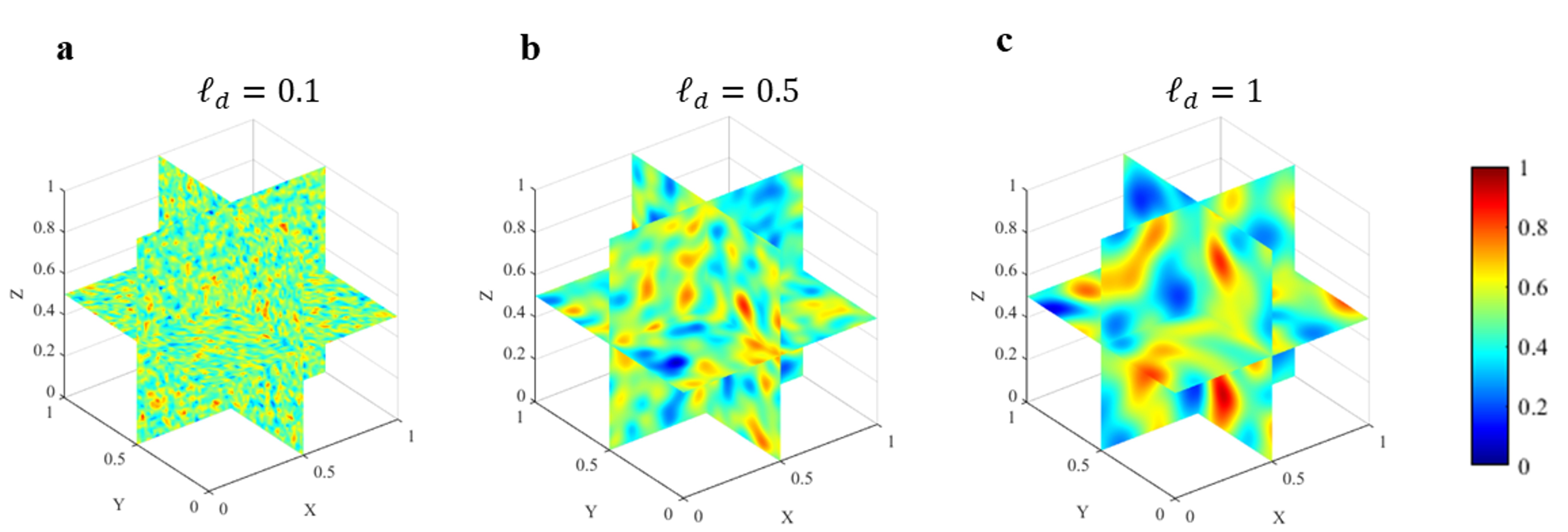}
    \caption{3D Gaussian Field with different \(\ell_d\): \textbf{a.} 3 slices-Gaussian Field when \(\ell_d=0.1\); \textbf{b.} 3 slices-Gaussian Field when \(\ell_d=0.5\); \textbf{c.} 3 slices-Gaussian Field when \(\ell_d=1\).}
    \label{fig:guass}
\end{figure}

In this subsection, we will elaborate on how to construct a training dataset that incorporates varying boundary conditions and source terms \textcolor{black}{based on GRF}, serving as the foundation for the experiments presented in Section \ref{sec4}. The training dataset consists of $N$ forcing functions, $f_j: \Omega \rightarrow \mathbb{R}$ and \textcolor{black}{boundary} conditions, $g_j: \partial\Omega \rightarrow \mathbb{R}$, and associated system responses, $u_j: \Omega \rightarrow \mathbb{R}$, which are solutions to the following equation:
\begin{equation}
    \mathcal{L} u_j=f_j, \quad \mathcal{D}\left(u_j, \Omega\right)=g_j,
\end{equation}
$f_j$ is the source term and $g_j$ is the constraint on the boundary. The training data comprises $N$ data pairs, where the forcing terms or the boundary conditions are drawn at random from a Gaussian process, $\mathcal{G} \mathcal{P}\left(0, \mathcal{K}\right)$, and $\mathcal{K}$ is the squared-exponential covariance kernel \cite{seeger2004gaussian} defined as

\begin{equation}
    \mathcal{K}(\boldsymbol{x}_i, \boldsymbol{x}_j) = \exp\left( -\frac{1}{2} \sum_{d=1}^3 \left(\frac{x_{id} - x_{jd}}{\ell_d}\right)^2 \right)\textcolor{black}{,} 
\end{equation}
where \( \boldsymbol{x}_i = (x_{i1}, x_{i2}, x_{i3}) \) and \( \boldsymbol{x}_j = (x_{j1}, x_{j2}, x_{j3}) \) are two points in three-dimensional space.
\( \ell_d \) denotes the length-scale parameter for each dimension \( d \). The parameter \(\ell_d > 0\) governs the correlation between the values of \(f \sim \mathcal{GP}(0, \mathcal{K})\) at points \(\boldsymbol{x}_i\) and \(\boldsymbol{x}_j\), where \(\boldsymbol{x}_i, \boldsymbol{x}_j \in \Omega\). As shown in Figure \ref{fig:guass}, a smaller value of \(\ell_d\) results in more oscillatory random functions. The diversity of the training set caused by \(\ell_d\) is essential for capturing the different modes of the operator \(\mathcal{L}\) and accurately learning the corresponding Green's function, as discussed in \cite{boulle2023learning}. In this work, \(\ell_d\) is selected within the range [0.1, 1], depending on the specific problem under consideration. \textcolor{black}{The GRF is implemented via a custom-developed Python script, which is included in our publicly available code repository.}
It is worth noting that we do not adopt a decoupled training strategy (i.e., setting one dataset \(g = 0\) to learn \(G\), and then another dataset \(f = 0\) to train \(\nabla G\)).
\textcolor{black}{In some cases, such configurations may lead to unsolvable boundary value problems.}

\subsection{Numerical quadrature and Loss function}
The computation of Eq. (\ref{eq3}) is central to this method, where Gaussian quadrature is employed to approximate it. Specifically, the computational domain \(\Omega\) is discretized into a mesh composed of unstructured tetrahedral elements in the interior and triangular surfaces on the boundaries. Specifically, the interior domain is divided into tetrahedral elements denoted as \(\mathcal{T} = \{T_l\}\), while the boundary is represented by triangular faces denoted as \(\mathcal{E}^{\text{bdry}} = \{E_m\}\). 

For the tetrahedral elements \(T_l\), a 4-point Gaussian quadrature is employed. For the triangular faces \(E_m\), a 3-point Gaussian quadrature is utilized. The formulas for determining the Gaussian quadrature points and their corresponding weights are provided as follows:

\[
\begin{aligned}
&\text{3-point quadrature rule on } E_m: \quad
\mathbf{\eta} = [\eta_{i1},\eta_{i2},\eta_{i3}]
= \begin{bmatrix}
\frac{2}{3} & \frac{1}{6} & \frac{1}{6} \\[6pt]
\frac{1}{6} & \frac{2}{3} & \frac{1}{6} \\[6pt]
\frac{1}{6} & \frac{1}{6} & \frac{2}{3} 
\end{bmatrix}, \quad
\mathbf{w} = \begin{bmatrix}
\frac{1}{3} \\[6pt]
\frac{1}{3} \\[6pt]
\frac{1}{3} 
\end{bmatrix}.\\
&\text{4-point quadrature rule on } T_l: \quad
\mathbf{\zeta} = [\zeta_{i1},\zeta_{i2},\zeta_{i3}]
= \begin{bmatrix}
0.58541020 & 0.13819660 & 0.13819660 \\[6pt]
0.13819660 & 0.58541020 & 0.13819660 \\[6pt]
0.13819660 & 0.13819660 & 0.58541020 \\[6pt]
0.13819660 & 0.13819660 & 0.13819660
\end{bmatrix}, \quad
\mathbf{w} = \begin{bmatrix}
\frac{1}{4} \\[6pt]
\frac{1}{4} \\[6pt]
\frac{1}{4} \\[6pt]
\frac{1}{4}
\end{bmatrix}.
\end{aligned}
\]
Here, each row of the matrix \(\mathbf{\eta}\) and \(\mathbf{\zeta}\), denoted as \(\mathbf{\eta}_i = [\eta_{i1}, \eta_{i2}, \eta_{i3}]\) and \(\mathbf{\zeta}_i = [\zeta_{i1}, \zeta_{i2}, \zeta_{i3}]\), represents the coordinates of the \(i\)-th Gaussian point.

\begin{figure}[H]
    \centering
    \includegraphics[width=0.9\linewidth]{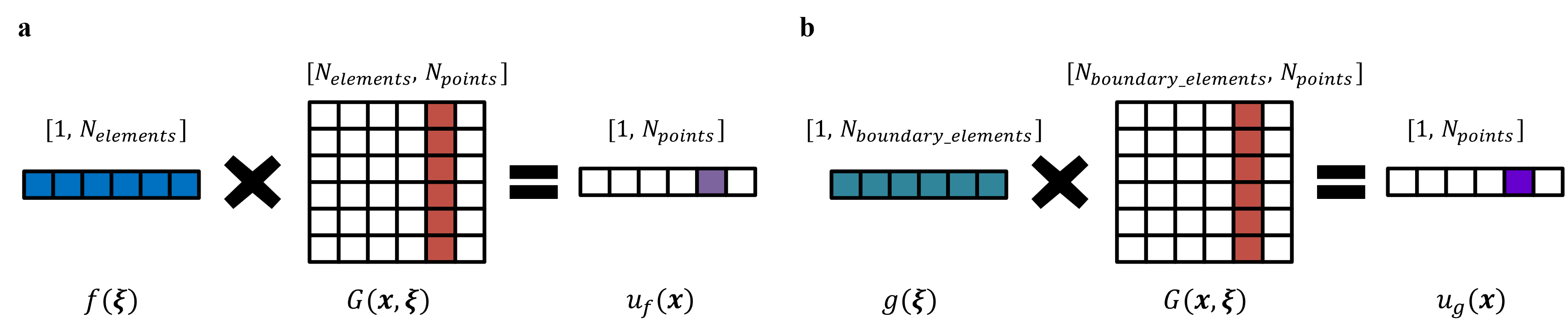}
    \caption{Illustration of numerical quadrature: (a) \( I_{\boldsymbol{\xi}, h}^{T_l}[f(\boldsymbol{\xi}) G(\boldsymbol{\xi}, \boldsymbol{x})] \) over the domain; (b) \( I_{\boldsymbol{\xi}, h}^{E_m}\left[g(\boldsymbol{\xi}) \nabla_{\boldsymbol{\xi}} G(\boldsymbol{\xi}, \boldsymbol{x})\right] \) over the boundary.}
    \label{fig:quadra}
\end{figure}

Then Eq. (\ref{eq3}) can be approximated as:
\begin{equation}
    \hat{u}(\boldsymbol{x}) \approx \sum_{T_i \in \mathcal{T}} I_{\boldsymbol{\xi}, h}^{T_l}[f(\boldsymbol{\xi}) G(\boldsymbol{\xi}, \boldsymbol{x})]-\sum_{E_m \in \mathcal{E}^{\text {bdry}}} I_{\boldsymbol{\xi}, h}^{E_m}\left[g(\boldsymbol{\xi}) a(\boldsymbol{\xi})\left(\nabla_{\boldsymbol{\xi}} G(\boldsymbol{\xi}, \boldsymbol{x}) \cdot \mathbf{n}_{\xi}\right)\right].
    \label{eq9}
\end{equation}
Here $I_{\boldsymbol{\xi}, h}^{T_l}[\cdot]$ denotes the numerical quadrature for evaluating $\int_{T_l} f(\boldsymbol{\xi}) G(\boldsymbol{\xi}, \boldsymbol{x}) d \boldsymbol{\xi}$ and $I_{\boldsymbol{\xi}, h}^{E_m}[\cdot]$ the quadrature for evaluating $\int_{E_m} g(\boldsymbol{\xi}) a(\boldsymbol{\xi})\left(\nabla_{\boldsymbol{\xi}} G(\boldsymbol{\xi}, \boldsymbol{x}) \cdot \mathbf{n}_{\xi}\right) d S(\boldsymbol{\xi})$, respectively. 
In Eq. (\ref{eq9}), \(\boldsymbol{\xi}\) in \(I_{\boldsymbol{\xi}, h}^{T_l}[\cdot]\) represents the Gaussian quadrature points from all elements in the mesh, with a total length of \(N_{\text{elements}} \times 4\). Similarly, \(\boldsymbol{\xi}\) in \(I_{\boldsymbol{\xi}, h}^{E_m}[\cdot]\) corresponds to the Gaussian quadrature points from all boundary elements of the mesh, with a total length of \(N_{\text{boundary\_elements}} \times 3\). As illustrated in Fig. \ref{fig:quadra}, the computational complexity for evaluating the numerical quadrature at a single point \(\boldsymbol{x}\) is \(O(N_{\text{elements}} + N_{\text{boundary\_elements}})\). For all points in the geometric domain, the overall complexity is \(O(N_{\text{points}} \times (N_{\text{elements}} + N_{\text{boundary\_elements}}))\). \textcolor{black}{Note that our framework can also be applied for the case without a mesh after modifying the quadrature scheme (e.g., using Monte Carlo integration).}

In our experiments, the loss function is defined as Eq. (\ref{eq10}), \textcolor{black}{where the $\hat{u}$ is calculated by convolution defined in Eq. (\ref{eq9}) and the exact $u$ is acquired by FEM method.} For a case involving 10,000 grid elements, the quadrature computation time \textcolor{black}{for calculating $\hat{u}$ for a sample $j$} is approximately 0.5 seconds, and the total training time is about 1 hour with 1.5 seconds \textcolor{black}{per training epoch (on an NVIDIA A100 GPU).} 
In prior studies \cite{teng2022learning, aldirany2024operator}, Green's functions were approximated within a data-free framework. 
\textcolor{black}{In \cite{teng2022learning}, the loss function is defined as \( Loss = \left(\mathcal{L} G\left(\boldsymbol{x}, \boldsymbol{\xi}\right) - \rho(\boldsymbol{x}, \boldsymbol{\xi})\right)^2 \), where \( \rho(\boldsymbol{x}, \boldsymbol{\xi}) \) represents a Gaussian density function employed to approximate the Dirac delta function.  
Similarly, in \cite{aldirany2024operator}, the loss function takes the form \( Loss = \left(\mathcal{L} \hat{u}\left(\boldsymbol{x}\right)-f(\boldsymbol{x})\right)^2 \), where \( \hat{u} \) denotes the solution obtained through the convolution between the approximated Green's function and the loadings.}
Although this approach eliminates the need for training data, its scalability is hindered by the computational cost of automatic differentiation, resulting in training times of up to 13 hours for 2D problems \textcolor{black}{(on a cluster of 16 NVIDIA RTX 2080 GPUs)}. In contrast, the proposed method substantially enhances training efficiency and accelerates convergence, enabling practical applications to more computationally demanding 3D problems.
Furthermore, \textcolor{black}{in \cite{teng2022learning}} the discrepancy between the Gaussian density function $\rho(\boldsymbol{x}, \boldsymbol{\xi})$ and the exact Dirac delta function $\delta(\boldsymbol{x}-\boldsymbol{\xi})$ is compounded through the superposition principle during the integration process for obtaining the solution. This cumulative amplification of the error ultimately degrades the accuracy of the final solution.
The enhanced efficiency and accuracy are key advantages of our approach.

\begin{equation}
    Loss = \sum^{N_{\text{samples}}}_{j=1} \sum^{N_{\text{points}}}_{i=1} (\hat{u}^j(\boldsymbol{x}_i)-u^j(\boldsymbol{x}_i))^2, \quad \boldsymbol{x}_i \in \Omega, \; i=1,...,N_{\text{points}}, \; j=1,...,N_{\text{samples}}.
    \label{eq10}
\end{equation}

\subsection{Binary structured neural network}
In this paper, we employ the binary structured neural network (BsNN) \cite{liu2024binary} as the fundamental component of both the Trunk Net and the Branch Net. \textcolor{black}{Compared with feed-forward neural networks (FNN), BsNN demonstrate superior efficiency and effectiveness in capturing the local features of solutions \cite{liu2024binary}. The rationale for selecting BsNN lies in the observation that the Green's functions possess singularities near the diagonal \cite{boulle2023learning}.}  

The BsNN is designed with the inspiration of “mixture of experts” (MoEs) model \cite{jacobs1991adaptive}, where the model comprises multiple independent expert networks. Each expert network specializes in solving a subproblem within a complex task, and their collective knowledge is combined to address the overall complex problem. The BsNN is similar to MoE, composed of multiple sub-networks, with each sub-network dedicated to learning a specific local feature of the solution and the collective knowledge gained by these sub-networks represents the complete set of solution features.
\textcolor{black}{In experiments demonstrated in Apendix A.2 (a), it is validated that BsNN achieve faster convergence compared to FNN. Moreover, for cases where FNN fail to converge (see Apendix A.2 (b)), BsNN successfully achieve normal convergence, further highlighting their efficiency.}

\begin{figure}[H]
    \centering
    \includegraphics[width=0.7\linewidth]{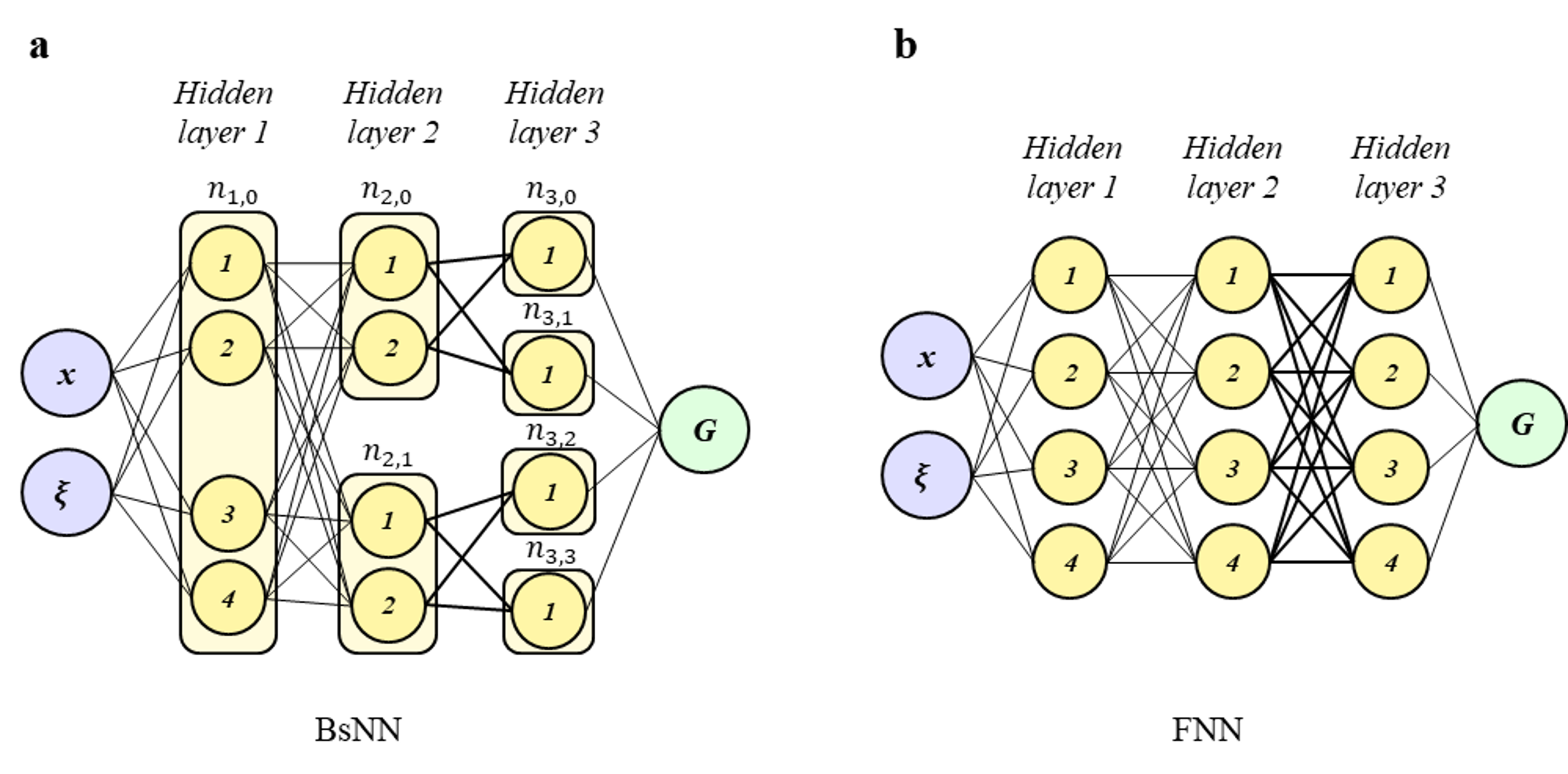}
    \caption{\textbf{a.} The structure of a BsNN consists of 3 hidden layers, each with 4 neurons (except the first and the last hidden layers, the neuron blocks in each hidden layer are fully connected to two neuron blocks in the next hidden layer). \textbf{b.} The structure of a FNN consists of 3 hidden layers, each with 4 neurons.}
    \label{fig:network}
\end{figure}

The general network structure of BsNNs is illustrated in Figure \ref{fig:network} \textbf{a}. Each $n_{i, j}$ in Figure \ref{fig:network} \textbf{a} contains one or more neurons, and such $n_{i, j}$ is referred to as a "neuron block". The black arrows indicate fully connected relationship between two neuron blocks, reflecting trainable weight parameters connecting every neuron pair from the two neural blocks. Within the final layer, the outputs of each neuron block are concatenated and fully connected to the output. This structure resembles a binary tree, where the neuron blocks in each hidden layer, except the first and the last, are fully connected to two neuron blocks in the next hidden layer. When the network possesses substantial depth, the parameter count of a BsNN is notably lower than that of an equivalently sized FNN featuring the same quantity of neurons. In the experiments conducted in this study, the number of parameters in the BsNN is approximately 1093, compared to 1393 in the FNN, representing a reduction of approximately 21\%.

In BsNN, $w_{i, j}, b_{i, j}$, and $\varsigma_{i, j}$ represent the weight, bias, and activation function, respectively, for the $j$-th branch of the $i$-th hidden layer $\left(j=0,1, \ldots, 2^i-1, i=0,1, \ldots, n-2\right)$. Correspondingly, $w_{n-1}, b_{n-1}$, and $\varsigma_{n-1}$ denote the weight, bias, and activation function for the final layer. Each neuron block within the same hidden layer contains the same number of neurons, and this quantity is known as the "block size" of that hidden layer. In this paper, the number of neuron blocks for the $i$-th layer with $1 \leq i<n$ is $2^{i-1}$, and for $1<i<n$, the block size in the $(i-1)$-th layer is twice that of the $i$-th layer.
With these notations, the forward propagation of the BsNN can be mathematically expressed as follows:
\begin{equation}
    o_{i, j}=\varsigma_{i-1, j}\left(w_{i-1, j} o_{i-1,\left\lfloor\frac{j}{2}\right\rfloor}+b_{i-1, j}\right), \quad i=1, \ldots,(n-1), j=0, \ldots,\left(2^{i-2}-1\right).
\end{equation}

The outputs $o_{n-1, j}$ (where $\left.j=0,1, \ldots,2^{n-2}-1\right)$ are concatenated along the last dimension to create a variable referred to as $o_{n-1}$. Subsequently, the final output of the BsNN is obtained as follows
\begin{equation}
    o_n=\varsigma_{n-1}\left(w_{n-1} o_{n-1}+b_{n-1}\right).
\end{equation}

\subsection{Parallel strategy for efficient calculation}
\begin{figure}[H]
    \centering
    \includegraphics[width=0.9\linewidth]{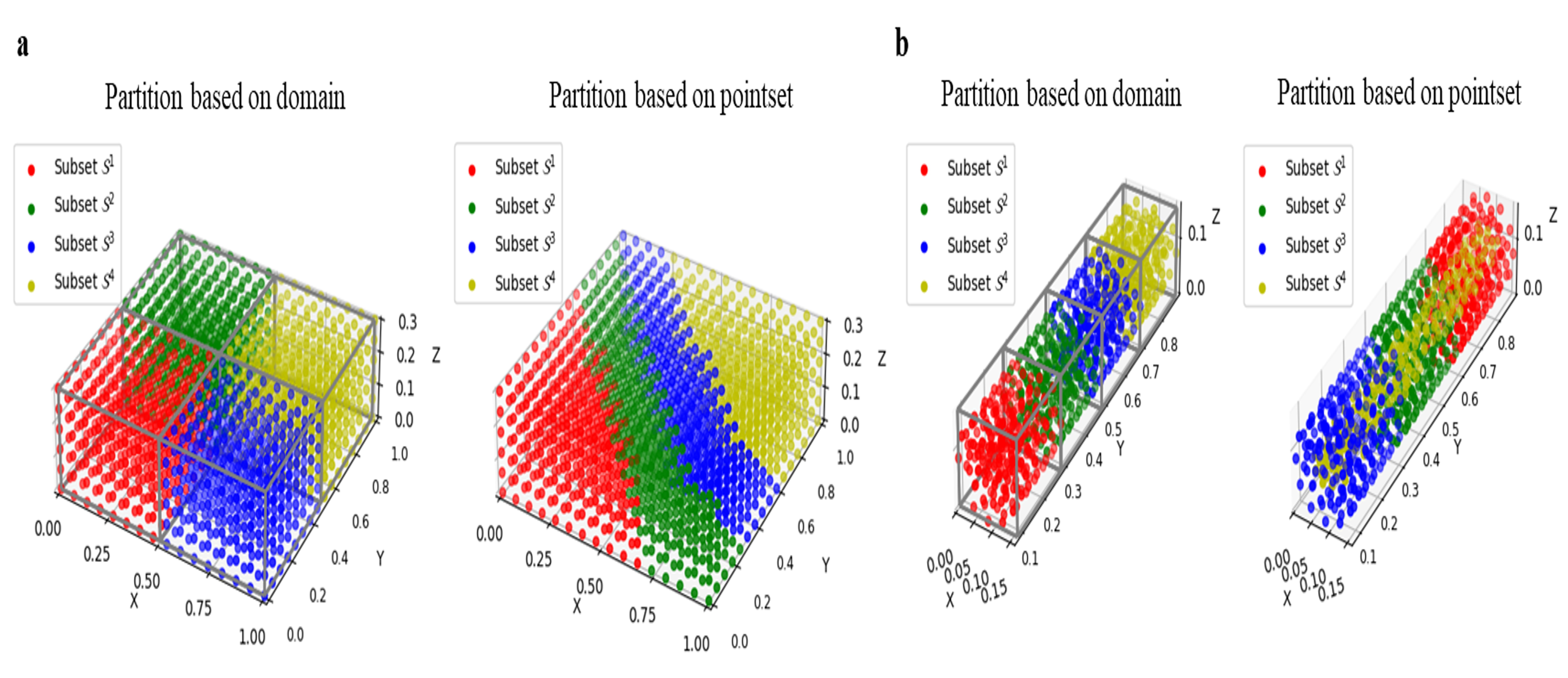}
    \caption{\textcolor{black}{Illustration of point set partition strategy: \textbf{a} partition strategy based on the domain of the regular plat, partition strategy based on the point-set of the regular plat; \textbf{b} partition strategy based on the domain of the pipe, partition strategy based on the point-set of the pipe.}}
    \label{fig:partition}
\end{figure}

When handling a large dataset \(\boldsymbol{x} \subset \Omega\), computing Eq. (\ref{eq9}) and performing backpropagation during training on a single GPU becomes computationally prohibitive. Since the Green's function corresponding to each point \(\boldsymbol{x} \in \Omega\) can be computed independently, the domain \(\Omega\), or equivalently the set of points \(\boldsymbol{x}\), can be partitioned to facilitate parallel computation. As illustrated in Fig. \ref{fig:partition}, two partitioning strategies can be employed: (a) dividing the computational domain \(\Omega\) into several subdomains of approximately equal size, with each subdomain consisting of a subset of points; or (b) directly splitting the entire set of points into several subsets of comparable size. For the computation of GONs, both strategies are largely equivalent in terms of parallelization efficiency. By assigning each partition to an independent GPU for parallel computation, the overall computational efficiency can be significantly improved.

We propose a highly parallelizable strategy for processing $\boldsymbol{x}$. Specifically, we partition the point set  
\(
\mathcal{S} = \left\{ (\boldsymbol{x}, \boldsymbol{\xi}) : \boldsymbol{x}, \boldsymbol{\xi} \in \Omega \right\}
\)  
into $K$ subsets of approximately equal size, denoted by  
\(
\mathcal{S}^k = \left\{ (\boldsymbol{x}, \boldsymbol{\xi}) : \boldsymbol{x} \in \mathcal{S}_{\boldsymbol{x}}^k, \boldsymbol{\xi} \in \mathcal{S}_{\boldsymbol{x}, \boldsymbol{\xi}} \right\},  
\)  
where $k = 1, \ldots, K$, and $n_\text{points}$ is the total number of points in $\mathcal{S}$. 
For each subset, a sub-training process is conducted to train the GON, gradually fine-tuning their parameters for each $\boldsymbol{x}$-block until all blocks have been processed. Using these partitioned samples, the complete training process is composed of $K$ sub-training tasks, each utilizing the data points in $\mathcal{S}_{\boldsymbol{x}}^k$. This approach naturally decomposes the training workload into smaller, independent subtasks, enabling efficient parallelization and implementation across multiple GPUs.

\subsection{Similarities and differences with DeepONet}
Drawing inspiration from DeepONet, we design the Trunk and Branch networks to enable hierarchical learning of the target operator.
It is worth noting that the use of Trunk Net and Branch Net in GON is purely for structural design and has no connection to the universal approximation theorem, where these terms were originally introduced.
Specifically, the Trunk Net within GON is designed to approximate the unknown Green’s functions \( G(\boldsymbol{x}, \boldsymbol{\xi}) \) of the system, while the Branch Net is utilized to approximate the auxiliary gradients of the Green’s function \( \nabla G(\boldsymbol{x}, \boldsymbol{\xi}) \). 
The introduction of the Branch Net is particularly beneficial in irregular geometric domains (such as finned tubes), where directly enforcing both \( G \) and its gradients \( \nabla G \) to satisfy the governing constraints through automatic differentiation and neural network backpropagation is challenging. This difficulty arises due to the inherent singularity of \( G \) and the poor convergence behavior often observed in automatic differentiation approaches for such complex domains. This is also the reason why we do not impose a compatibility penalty term on 
\( \nabla G \) in the loss function. By explicitly learning \( \nabla G \) with the Branch Net, we enhance the solution accuracy. For regular geometric domains (such as cubes), where the learning difficulty is lower and sufficient accuracy can be achieved without additional modifications, the Branch Net can be omitted.
A detailed discussion and validation of the effectiveness of the Branch Net are provided in Appendix A.3.

\begin{figure}[H]
    \centering
    \includegraphics[width=0.9\linewidth]{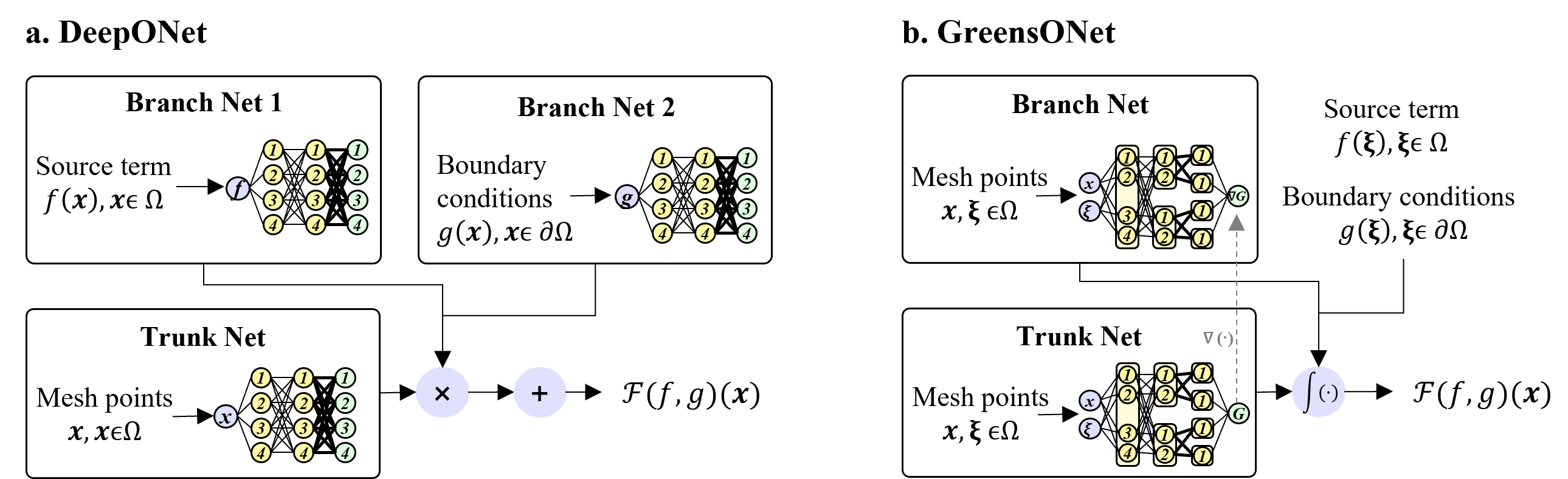}
    \caption{Illustration of of (a) structure of DeepONet and (b) structure of GON.}
    \label{fig:comparision}
\end{figure}

To better illustrate the similarities and differences between our method and DeepONet, we depict their respective structures in Fig.~\ref{fig:comparision}. DeepONet consists of two Branch Nets and one Trunk Net. The Trunk Net takes coordinate information as input, while the two Branch Nets encode the source term and boundary conditions, respectively. The outputs of the Trunk Net and the Branch Nets are combined using the Hadamard product, followed by a summation to yield the target physical quantity. It is worth noting that for multi-dimensional target quantities, the summation step is omitted.  

In contrast, GON requires only one Trunk Net and one Branch Net. Both the Trunk Net and the Branch Net take coordinate information as input, with their outputs representing the desired Green's function and its gradient. Notably, the gradient of the Green's function can alternatively be computed via automatic differentiation, in which case the Branch Net can be omitted. The target physical quantity is then obtained through numerical integration of the outputs. For multi-dimensional physical quantities, Green's function matrices are learned by employing multiple pairs of Trunk Nets and Branch Nets. These matrices are subsequently integrated with the boundary conditions and source terms to compute the desired quantities. In summary, while DeepONet relies on multiple Branch Nets and a Trunk Net with Hadamard product operations to approximate target quantities, GON adopts a more streamlined approach by leveraging numerical integration of Green's functions, offering greater flexibility and efficiency.

\textcolor{black}{The algorithm of the GON can be summarized as Algorithm 1.}

\begin{algorithm}
\caption{Solving PDEs by GON and acquiring Green's function}
\begin{algorithmic}[1]
\STATE \textbf{Input:} $\mathcal{L}(\cdot), f(\boldsymbol{x})$ and $g(\boldsymbol{x})$, the mesh $\mathcal{T}_q$, and an interior vertex $\boldsymbol{x} \in \Omega$
\STATE \textbf{Output:} The PDE solution at $\boldsymbol{x}: u(\boldsymbol{x})$
\STATE Generate the quadrature points and quadrature weights for each element.
\STATE Calculate the normal and volume for each element in the domain.
\STATE Calculate the area for each element on the boundary.
\IF{need parallel}
    \STATE Apply the domain partition to divide $\mathcal{S}_{\boldsymbol{x}, \boldsymbol{\xi}}$ into $K$ blocks ($K>1$).
\ELSE
    \STATE $K$=1.
\ENDIF
 
\FOR{1, $N_{samples}$}
    \FOR{$k$=1,...,K}
        \FOR{1,...,$N_{epoches}$}
            \STATE Feed all points in $\mathcal{S}_{\boldsymbol{x}, \boldsymbol{\xi}}^k$ into the Branch Net and Trunk Net.
            \STATE Acquire $G\left(\boldsymbol{x}, \boldsymbol{\xi}; \boldsymbol{\Theta}\right)$ and $\nabla$ $G\left(\boldsymbol{x}, \boldsymbol{\xi}; \boldsymbol{\Theta}\right)$.
            \STATE Calculate $\hat{u}(\boldsymbol{x})$ by integration defined in Eq. (\ref{eq9}).
            \STATE Calculate error of $(\hat{u}(\boldsymbol{x})-u(\boldsymbol{x}))^2$.
            \STATE Optimize parameters $\boldsymbol{\Theta}$ of the Branch Nets and Trunk Nets to minimize error of $(\hat{u}(\boldsymbol{x})-u(\boldsymbol{x}))^2$.
        \ENDFOR
    \ENDFOR
\ENDFOR

\STATE Return well-trained Greens' function.
\end{algorithmic}
\end{algorithm}

\section{Experiments and results \label{sec4}}
In this section, we evaluate the performance of the proposed GON framework for approximating Green’s functions and its application to efficiently solving three classical PDEs using Algorithm 1. The investigated cases include:  

\textcolor{black}{Case 0. \textit{2D Poisson Equation}:  
    This case is specifically designed to enable a fair comparison with existing Green’s function-based methods, which have primarily focused on 2D problems. To ensure consistency, we adopt a 2D setting as a baseline for performance evaluation.}

Case 1. \textit{3D Steady Heat Conduction Equation}:  
   This case examines scenarios involving a finned tube under varying boundary conditions.  

Case 2. \textit{3D Heterogeneous Reaction-Diffusion Equations}:  
   Two scenarios are considered: homogeneous diffusion on a flat plate and heterogeneous mixing between two substances within a micro-pipe.  

Case 3. \textit{3D Stokes Equations}:  
   The analysis focuses on the effects of source terms on fluid flow. In contrast to the previous cases, this investigation involves constructing a Green’s function matrix to solve for multidimensional velocity variables.  

\textcolor{black}{For each case, we provide detailed hyperparameter settings for GON and the corresponding baseline models in the respective subsections. In Case 0, the comparison focuses on existing Green’s function-based methods, which are predominantly limited to 2D problems. In contrast, Cases 1–3 emphasize comparisons with baseline models applicable to 3D problems (FNO, DeepONet, PI-DeepONet, and PINN), aiming to demonstrate the generality of our approach.}
All experiments reported in this study were performed on a remote server running Ubuntu 22.04 LTS, equipped with an \textcolor{black}{Intel® Xeon® Platinum 8380 processor (2.30GHz)} and an NVIDIA A100 GPU with 80GB of HBM2 memory.  

{\color{black}
\subsection{Case 0: Comparison with existing Green’s function-based methods \label{sec:4.0}}
Our approach builds on prior work that employs Green’s function approximation for solving PDEs and extends it by developing a more general framework that is boundary-invariant and source-term-invariant for 3D bounded domains. In this section, we compare our approach with existing methods that utilize Green's functions to solve linear equations \cite{boulle2022data, teng2022learning, aldirany2024operator}. A detailed comparison of these methods is provided in Table 1. Considering the approaches proposed in \cite{boulle2022data} and \cite{aldirany2024operator} are unable to handle varying boundary conditions, while \cite{boulle2022data}, \cite{teng2022learning}, and \cite{aldirany2024operator} have been applied exclusively to 1D and 2D cases. To ensure a fair evaluation, we conduct comparative experiments on a 2D Poisson problem with invariant boundary conditions:

\begin{equation}
    \left\{\begin{aligned}
    \nabla^2 u(\boldsymbol{x})=f(\boldsymbol{x}), \quad & \boldsymbol{x} \in \Omega \\
    u(\boldsymbol{x})=0. \quad & \boldsymbol{x} \in \partial \Omega
    \end{aligned}\right.
    \label{eq:poisson}
\end{equation}

The exact solutions are defined as:

\begin{equation}
    u(x, y) = C sin(2\pi x)sin(2 \pi y), \quad x, y \in \Omega.
\end{equation}
where $\Omega=[-1,1]\times[-1,1]$ and $C$ is a constant with varied value for different cases. 

\begin{table}[H]
    \centering
    {\color{black}
    \caption{A comparative analysis of Green's function based methods, highlighting key aspects of type of Neural Networks (NN), Loss function, Integration method, Boundary Conditions (BC), Source term ($f$), and Problem Dimensionality. Here, FNN represents feed-forward neural networks.}
    \begin{tabular}{c|c|c|c|c|c|c}
        \toprule
         Literature &  NN & Loss function & Integration method & BC & $f$ & Dimensionality \\
         \midrule
        \cite{boulle2022data} Boulle et al. 2022  & Rational NN & Data constraint & Monte-Carlo integration & $\times$ & \checkmark & 1D \& 2D\\
        \hline
        \cite{teng2022learning} Teng et al. 2022  & FNN & PDE constraint & Gauss integration & \checkmark & \checkmark & 1D \& 2D\\
        \hline
        \cite{aldirany2024operator} Aldirany et al. 2024  & FNN & PDE constraint & Monte Carlo integration & $\times$ & \checkmark & 1D \& 2D\\
        \hline
        Ours & BsNN & Data constraint & Gauss integration & \textbf{\checkmark} & \textbf{\checkmark} & \textbf{3D}\\
        \bottomrule
    \end{tabular}
    }
    \label{tab:method_compare}
\end{table}
}

\begin{table}[H]
    \centering
    {\color{black}
    \caption{Comparison with existing Green’s function-based methods for 2D poisson case.}
    \renewcommand{\arraystretch}{1.25}
    \begin{tabular}{c|c|c|c}
    \toprule
       Model & Training error & Testing error & Inference time  \\
       \midrule
        \cite{boulle2022data} Boulle et al. 2022 & $2.51\times 10^{-3}$ & $2.56\times 10^{-2}$ & 1.65 s\\
       \hline
       \cite{teng2022learning} Teng et al. 2022 & $2.63\times 10^{-2}$ & $1.05\times 10^{-1}$ & 2 s\\
       \hline
       \cite{aldirany2024operator} Aldirany et al. 2024 & $9.75\times 10^{-3}$ & $3.91\times 10^{-2}$ & 1.32 s\\
       \hline
       \textbf{Ours} & $\mathbf{9.38\times 10^{-5}}$ & $\mathbf{3.75\times 10^{-4}}$ & 1.36 s\\
    \bottomrule
    \end{tabular}
    }
    \label{tab:compare_green}
\end{table}

\begin{figure}[H]
    \centering
    \includegraphics[width=0.9\linewidth]{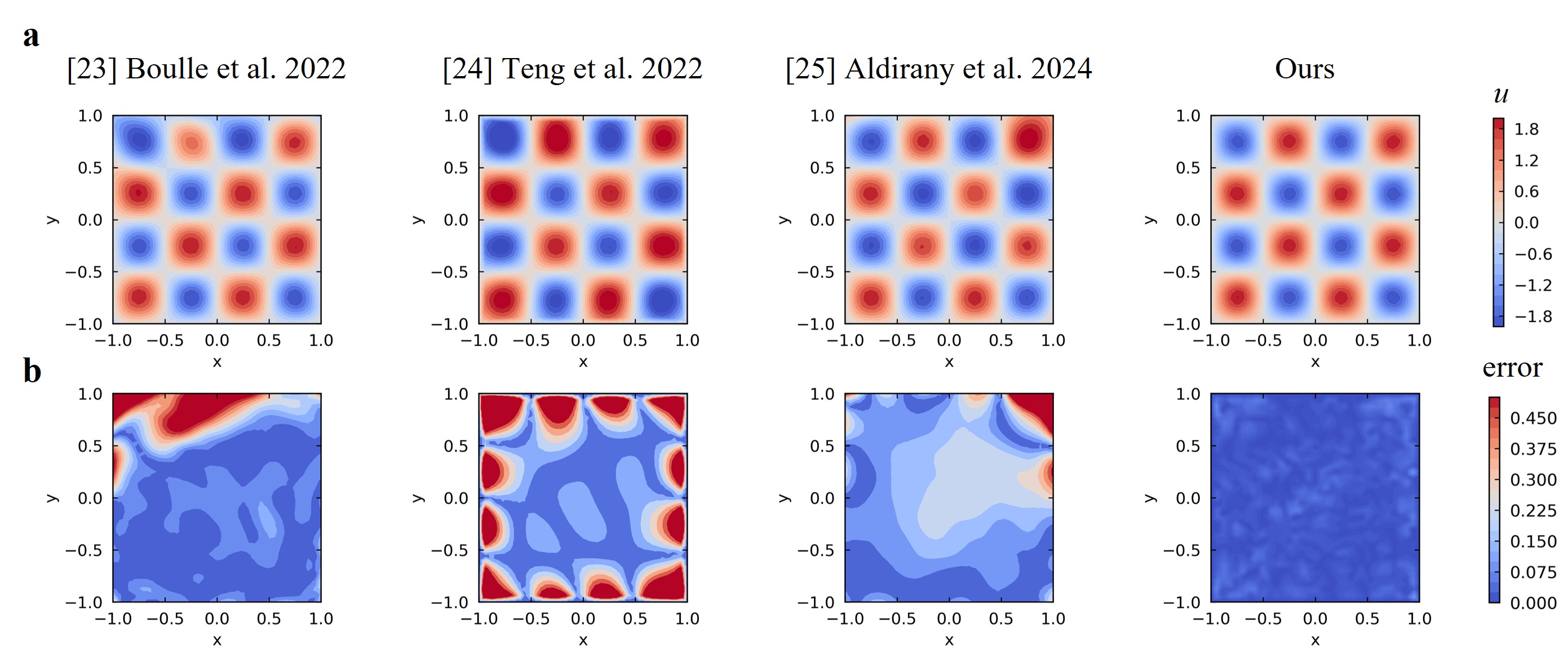}
    \caption{\textcolor{black}{Comparison with existing Green’s function-based methods: (a) Results acquired by methods in \cite{boulle2022data}, \cite{teng2022learning}, \cite{aldirany2024operator} and our method respectively; (b) Difference between exact solutions and the acquired solutions ($|\hat{u}-u|$).}}
    \label{fig:2d_poisson_case}
\end{figure}

{\color{black}
The hyperparameter are \textcolor{black}{kept} the same for fairness (Epochs =1000, Learning rate = 0.001, Layers = [4,12,12,12,1]).
The performance for different existing Green's function-based methods are demonstrated in  Table 2 and Figure \ref{fig:2d_poisson_case}.

In comparison with the method presented in \cite{boulle2022data}, as shown in Table 2, our approach \textcolor{black}{achieves lower errors} on both the training and testing sets.
\textcolor{black}{One contributing factor} is the integration scheme employed. In \cite{boulle2022data}, Monte Carlo integration is used to evaluate the convolution between the source term and the Green’s function. However, Monte Carlo methods converge at a rate of \( O(N^{-1/2}) \), requiring a large number of samples \( N \) to achieve acceptable accuracy, particularly in 2D problems \cite{liu2001monte, gil2007numerical}. In contrast, our method adopts Gaussian quadrature, which offers a convergence rate of \( O(N^{-k}) \), where \( k \) depends on the order of the quadrature rule. \textcolor{black}{In addition, theoretical singularities of Green's function are approximated by smooth surrogates in the learning process, allowing Gaussian quadrature to be applied effectively. In such settings, Gaussian quadrature can deliver higher accuracy with fewer points compared to Monte Carlo methods.}

In comparison with the methods proposed in \cite{teng2022learning} and \cite{aldirany2024operator}, our approach also achieves better performance on both the training and testing sets. 
A key distinction from \cite{teng2022learning} lies in the design of the loss function. Teng et al. approximate the Dirac delta function \( \delta(\boldsymbol{x}, \boldsymbol{\xi}) \) using a Gaussian function \( \rho(\boldsymbol{x}, \boldsymbol{\xi}) \), and extract the Green’s function by minimizing the PDE residual \( (\mathcal{L} G(\boldsymbol{x}, \boldsymbol{\xi}) - \rho(\boldsymbol{x}, \boldsymbol{\xi}))^2 \). However, this Gaussian approximation may hinder the network’s ability to accurately capture the steep gradient near the singularity, potentially degrading the solution accuracy in both training and inference phases.
In contrast, the method in \cite{aldirany2024operator} approximates the Green’s function indirectly by enforcing the PDE constraint \( (\mathcal{L} u(\boldsymbol{x}) - f(\boldsymbol{x}))^2 \), thereby reducing reliance on labeled data. While physics-informed approach enhances robustness, they are often limited by convergence difficulties \cite{sheng2021pfnn, lin2021binet, sukumar2022exact}.
}

\subsection{Case 1: Steady heat conduction equations \label{sec:4.1}}
Heat conduction equations play a fundamental role in mathematical physics and engineering, governing a wide range of phenomena such as metal smelting and heat dissipation in electronic components. In this study, we focus on the efficient solutions of the steady heat conduction equations under varying boundary conditions (ref. Eq \ref{eq:poisson}), which is critical for capturing diverse physical scenarios. A classical heat transfer case on the finned tube is considered. The physical condition setting and results are demonstrated below.

\begin{equation}
    \left\{\begin{aligned}
    \nabla^2 u(\boldsymbol{x})=Q(\boldsymbol{x}), \quad & \boldsymbol{x} \in \Omega \\
    u(\boldsymbol{x})=g(\boldsymbol{x}). \quad & \boldsymbol{x} \in \partial \Omega
    \end{aligned}\right.
    \label{eq:poisson}
\end{equation}

\begin{figure}[H]
    \centering
    \includegraphics[width=0.9\linewidth]{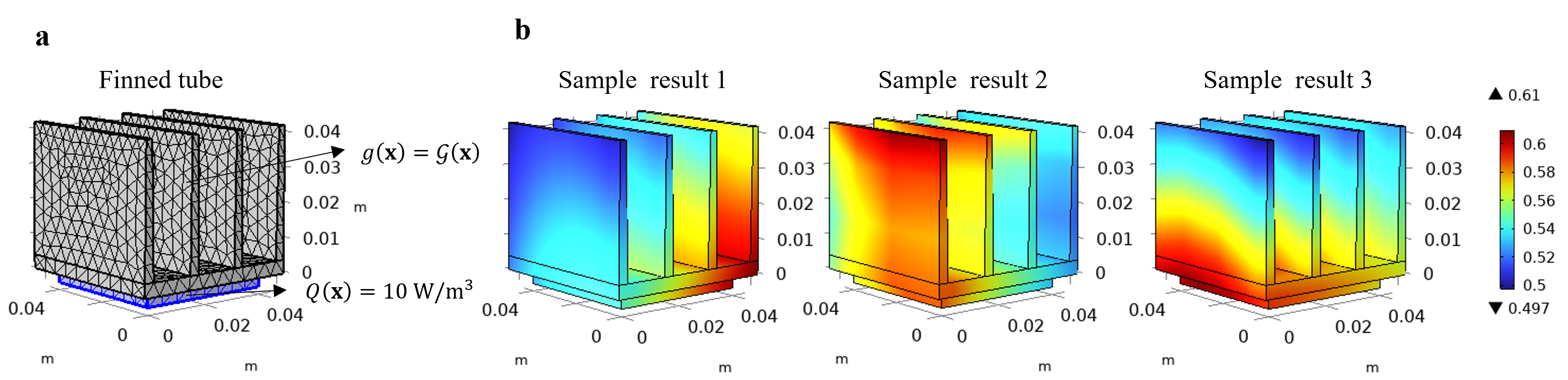}
    \caption{(a) The simulation setting of the steady heat conduction case on finned tube; (b) sample results with generated GRF conditions.}
    \label{fig:poisson_case}
\end{figure}

As illustrated in Figure \ref{fig:poisson_case}, the source term \( Q(\boldsymbol{x}) \) is set to 10 on the heated bottom plate, while the Dirichlet boundary condition \( g(\boldsymbol{x}) \) is derived from a GRF with a wavelength parameter \( \lambda = 1 \).
For the finned tube geometry, the computational mesh consists of 4427 tetrahedral cells and 1478 vertices within the domain, along with 2968 triangular faces on the boundary. The GRF boundary condition is applied to the top surface. In this study, we generated 100 datasets, with sample results illustrated in Figure \ref{fig:poisson_case} \textbf{b}. Among these, \textcolor{black}{70} datasets were used for training and \textcolor{black}{30} for testing. The datasets were generated using \text{COMSOL Multiphysics®}. \textcolor{black}{Since the exact solutions are not available, we use the numerical solution obtained on the finest mesh (with an average mesh size of $3.19 \times 10^{-4}\,\mathrm{m}$) as a surrogate reference. The approximate error, measured against this reference, is $5.11 \times 10^{-5}$ in the $L^2$ norm for Case 1.  If higher accuracy is desired, one can construct the dataset using even finer mesh resolutions during sample generation.}

\begin{figure}[H]
    \centering
    \includegraphics[width=0.75\linewidth]{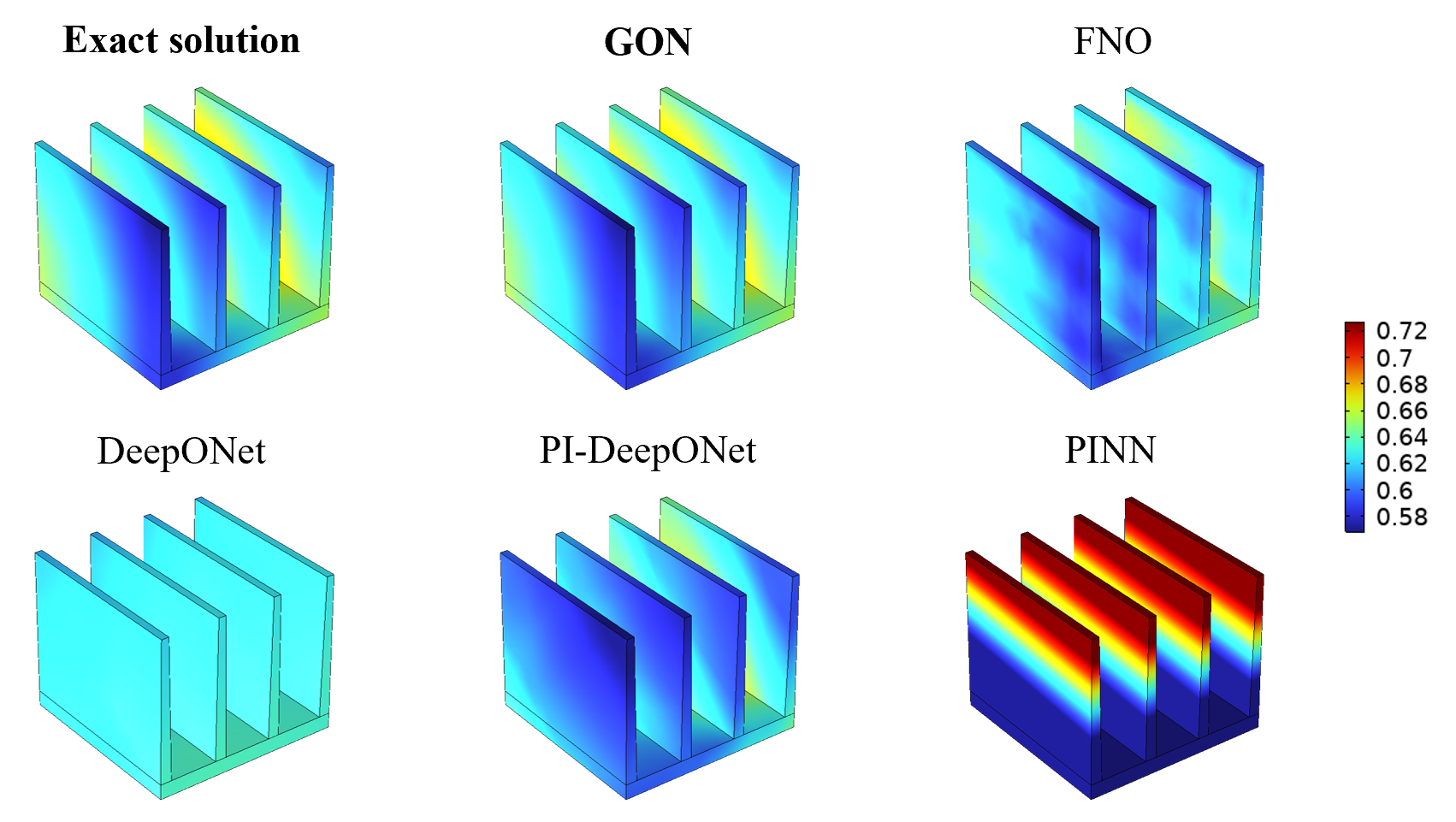}
    \caption{\textcolor{black}{Case study for steady heat conduction equation: temperature distribution from exact solution, and inference of GON, FNO, DeepONet, PI-DeepONet, PINN  (from left column to right column, from first row to second row).}}
    \label{fig:poisson_res2}
\end{figure}

For validation, we compared the GON framework against four well-known models: PINN \cite{raissi2019physics}, DeepONet \cite{lu2021learning}, PI-DeepONet \cite{wang2021learning}, and FNO \cite{li2020fourier}. A detailed introduction to PINN, DeepONet, PI-DeepONet, and FNO can be found in Section \ref{sec1}. Table \ref{tab:poisson2} and Figure \ref{fig:poisson_res2} present a comprehensive comparison between GON and these baseline models on the finned tube. To ensure a fair evaluation, we used identical hyperparameters across all models, including learning rate and optimizer settings. Additionally, given the variations in convergence rates among the different networks, we selected a sufficient number of iterations for each model to ensure convergence. 
To better assess the performance of each network, we kept the number of layers as consistent as possible across all \textcolor{black}{MLP-based models}.

As shown in Table \ref{tab:poisson2}, GON achieves the lowest testing error compared to PINN, DeepONet, PI-DeepONet, and FNO. It is important to note that FNO can only handle cases with regular geometries, requiring the finned tube geometry to be interpolated onto a structured grid for input into the FNO, with the output subsequently interpolated back to the original geometry. 
Figure \ref{fig:poisson_res2} further illustrates a primary limitation of FNO: although FNO captures the overall trend of the temperature distribution, it fails to represent fine-scale details accurately, especially when compared to GON.
This underscores the importance of incorporating physics-informed priors to enhance the generalization capabilities of neural networks. 

\begin{table}[H]
    \centering
    \caption{The hyper-parameters and performance of different baseline models on case of finned tube.}
    \renewcommand{\arraystretch}{1.25}
    \begin{tabular}{c|c|c|l|c|c}
    \toprule
    Model & Epochs & Learning rate & Layers & Training error & Testing error\\
    \midrule 
    GON & 2000 & 0.001 & [6, 12, 12, 12, 1] & \textcolor{black}{$\mathbf{2.18\times10^{-5}}$} & \textcolor{black}{$\mathbf{2.64\times10^{-5}}$} \\
    \hline
    FNO & \textcolor{black}{1500} & 0.001 & \textcolor{black}{[2, 3, 3, 3, 3, 3, 1]} & \textcolor{black}{$1.01 \times 10^{-4}$} & \textcolor{black}{$6.32 \times 10^{-4}$} \\
    \hline
    \multirow{3}{*}{DeepONet} & \multirow{3}{*}{2000} & \multirow{3}{*}{0.001} & Trunk Net: [3, 12, 12, 12, 1] & \multirow{3}{*}{\textcolor{black}{$8.14\times10^{-4}$}} & \multirow{3}{*}{\textcolor{black}{$1.37\times10^{-3}$}} \\
        & &  &  Branch Net 1: [$N_{points}$, 12, 12, 12, 1] & \\
        & &  &  Branch Net 2: [$N_{points}$, 12, 12, 12, 1]   & \\
    \hline
    \multirow{3}{*}{PI-DeepONet} & \multirow{3}{*}{2000} & \multirow{3}{*}{0.001} & Trunk Net: [3, 12, 12, 12, 1]  & \multirow{3}{*}{\textcolor{black}{$4.14\times10^{-4}$}} & \multirow{3}{*}{\textcolor{black}{$1.78\times10^{-3}$}} \\
        & &  & Branch Net 1: [$N_{points}$, 12, 12, 12, 1] &   \\
        & &  &  Branch Net 2: [$N_{points}$, 12, 12, 12, 1]   & \\
    \hline
     PINN  & 10000 & 0.001 & [3, 12, 12, 12, 1] & - & \textcolor{black}{$2.56\times10^{-2}$} \\
    \bottomrule
    \end{tabular}
    \label{tab:poisson2}
\end{table}

In this case, DeepONet consists of a Trunk Net with layers [3, 12, 12, 12, 1] to process position information \(\mathbf{x}\), along with two Branch Net: Branch Network 1 for handling the force term \(Q\) and Branch Network 2 for managing boundary conditions \(g\), as detailed in Table \ref{tab:poisson2}. In this setup, \(N_{points}\) denotes the total number of points in the domain, which is 1478 for the finned tube. 
As shown in Table \ref{tab:poisson2}, DeepONet achieves an \(L_2\) error of \(8.14 \times 10^{-4}\) on the test set. PI-DeepONet, which incorporates the same Trunk Net and additional Branch Net (Branch Net 1 and Branch Net 2), results in a lower \(L_2\) error of \(4.14 \times 10^{-4}\) compared to DeepONet. Furthermore, the distribution of isotherms in PI-DeepONet better aligns with the exact solution than that of DeepONet. This improvement may be attributed to the incorporation of the PDE constraints in the loss function.

PINN, which uses a similar five-layer structure [3, 12, 12, 12, 1] as GON, achieved an \(L_2\) error of \(2.56 \times 10^{-2}\) after 10000 epochs. Boundary and PDE constraints were incorporated into the loss function as soft penalties. The PDE loss is approximately \(1.47 \times 10^{-1}\) and could not decrease further. This suggests that, for the random GRF boundary conditions, PINN  may not be able to simultaneously satisfy both the random boundary condition and the PDE constraints using the Adam optimization algorithm.

Across all visualizations, GON is the only model capable of accurately capturing the full range of the temperature field, as indicated by the colorbar. Other baseline models exhibit varying degrees of deviation in this regard.

This case demonstrates the boundary condition adaptability of our method, highlighting its robustness and versatility in solving the heat conduction equation on both regular and irregular computational domains with varying boundary conditions.

\subsection{Case 2: Heterogeneous reaction-diffusion equations \label{sec:4.2}}
Reaction-diffusion equations play a crucial role in modeling complex spatial and temporal patterns in biological, chemical, and physical systems. 
In this work, we focus on steady heterogeneous reaction-diffusion equations under varying source terms. The study explores two classical cases: one in \textbf{(a)} a flat plate geometry and the other \textbf{(b)} in a pipe configuration. These models serve as representative examples to investigate the impact of spatially varying sources on the reaction-diffusion dynamics in different domains, with heterogeneous diffusion coefficients. The physical condition setting and results are demonstrated below.

\begin{equation}
    \left\{\begin{aligned}
    -\nabla \cdot(a(\boldsymbol{x}) \nabla u(\boldsymbol{x}))+r(\boldsymbol{x}) u(\boldsymbol{x}) = f(\boldsymbol{x}), \quad & \boldsymbol{x} \in \Omega \\
    u(\boldsymbol{x}) = g(\boldsymbol{x}). \quad & \boldsymbol{x} \in \partial \Omega
    \end{aligned}\right.
    \label{eq:diffusion}
\end{equation}

\begin{figure}[H]
    \centering
    \includegraphics[width=0.8\linewidth]{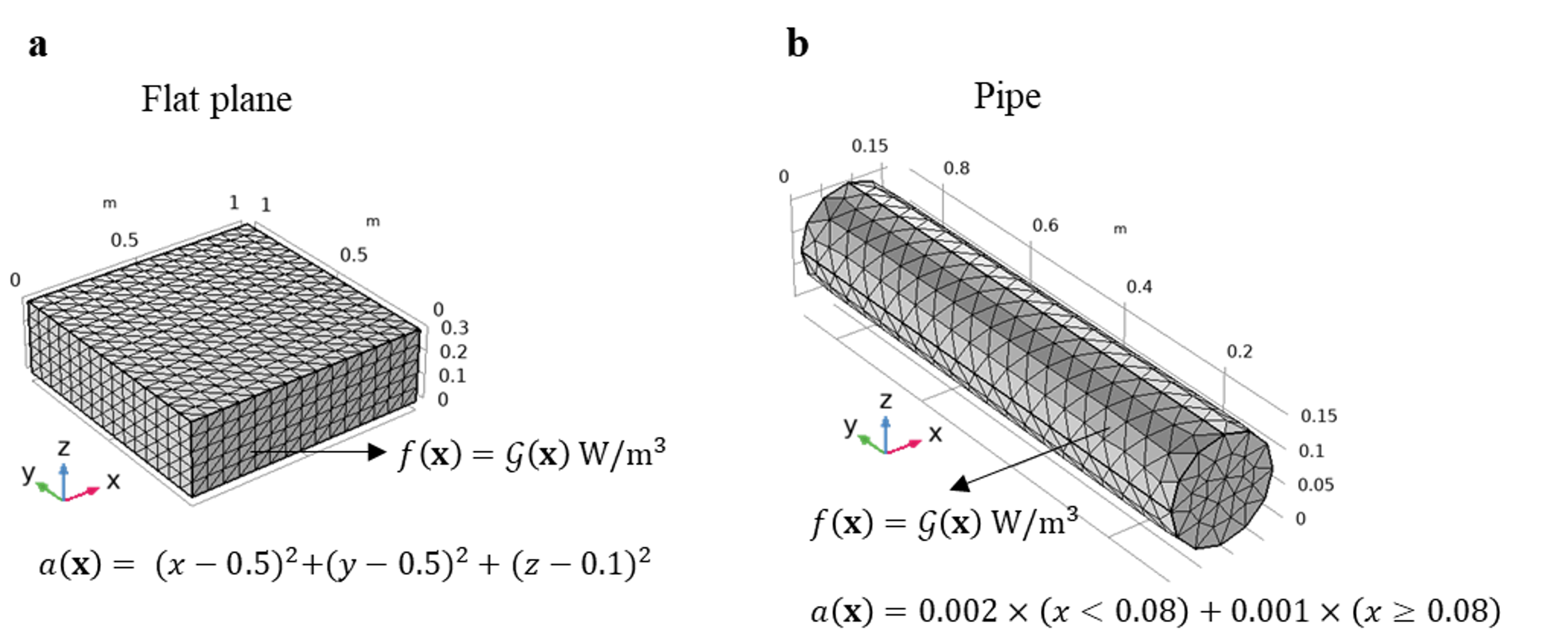}
    \caption{\textbf{a.} The simulation setting of the steady heterogeneous reaction-diffusion case on flat plane. \textbf{b.} The simulation setting of the steady heterogeneous reaction-diffusion case on pipe.}
    \label{fig:case_diffusion}
\end{figure}

\textbf{(a) Reaction-diffusion equations case on a flat plate.} 
As shown in Figure \ref{fig:case_diffusion} \textbf{(a)}, \textcolor{black}{the computational domain is a cylinder with a radius of 0.08 m along the $x$-axis and a height of 0.8 m along the $y$-axis.} The diffusion coefficient is set as $a(\mathbf{x})=(x-0.5)^2+(y-0.5)^2+(z-0.1)^2$, and the force term \(f(\boldsymbol{x})\) is generated by the GRF with a wavelength parameter \(\lambda = 1\). This setup is designed to simulate a natural phenomenon where diffusion is faster at the edges and slower in the center. For the flat plane, the mesh consists of 6750 equally sized tetrahedral cells and 1536 vertices within the domain, along with 1500 equally sized triangular faces on the boundary. In this study, we generated 100 datasets, of which 70 datasets were used for training and 30 datasets for testing.
The datasets were generated using \text{COMSOL Multiphysics®}. \textcolor{black}{The reference solution is obtained on a mesh with average size $7.20 \times 10^{-3}\,\mathrm{m}$, yielding an estimated $L^2$ norm error of $7.11 \times 10^{-2}$.}

\begin{figure}[H]
    \centering
    \includegraphics[width=0.85\linewidth]{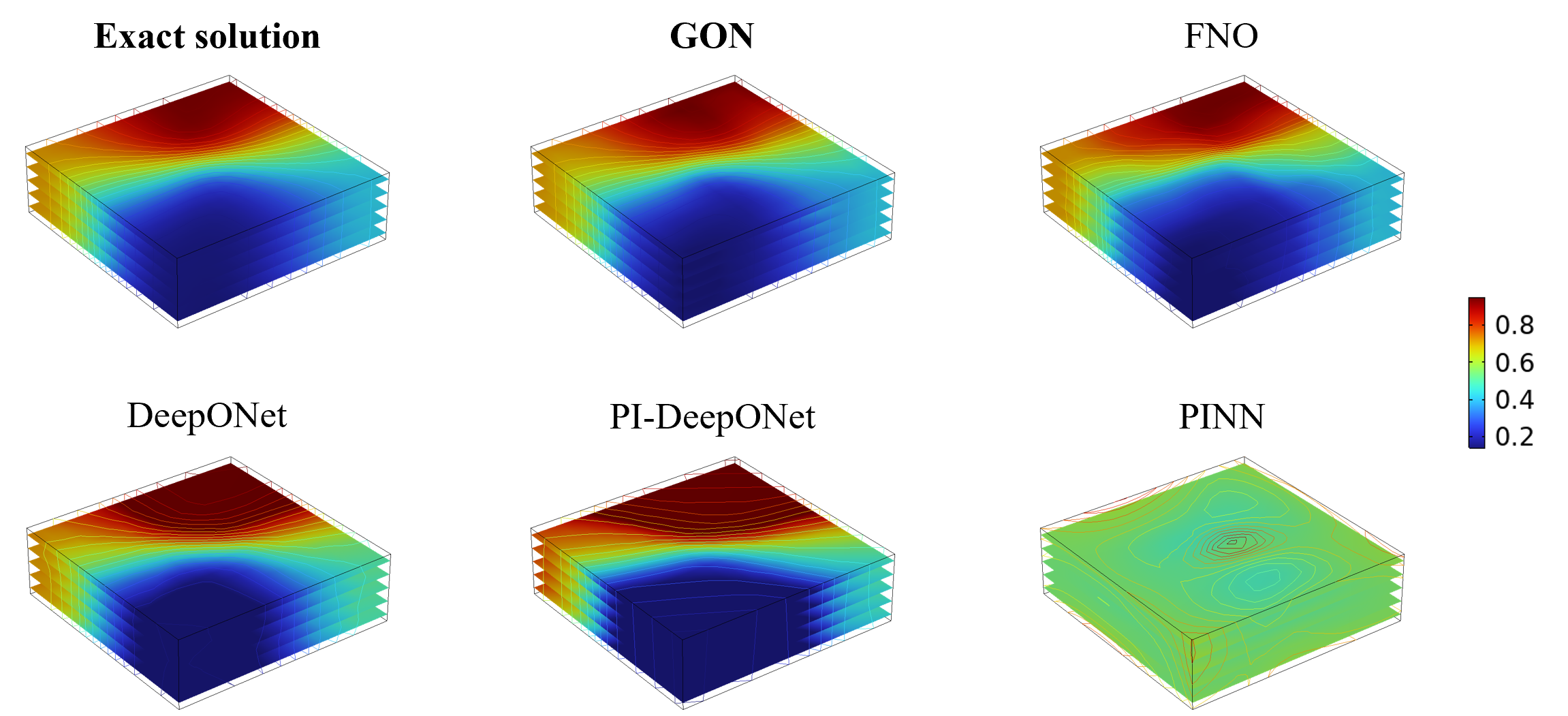}
    \caption{Case study for steady heterogeneous reaction-diffusion on flat plane: physical distribution and isotherm distribution from exact solution, and inference of GON, FNO, DeepONet, PI-DeepONet, PINN  (from left column to right column, from first row to second row).}
    \label{fig:diffusion}
\end{figure}

Consistent with the observations in Section 4.1, Table \ref{tab:diffusion} and Figure \ref{fig:diffusion} shows that DeepONet and PI-DeepONet perform worse than FNO and GON, with testing errors of \(3.68 \times 10^{-3}\) and \(4.02 \times 10^{-3}\), respectively. Furthermore, PINN encounters convergence difficulties, resulting in a significantly higher error of \(6.54 \times 10^{-2}\). These results validate the limitations of MLP-based methods compared to CNN-based frameworks, such as FNO, in effectively learning from high-dimensional data, as stated in \cite{lu2021learning}. Notably, although GON is also based on an MLP architecture, it surpasses the performance of FNO, underscoring the importance of incorporating physics-informed priors to enhance model effectiveness.

It is important to highlight that, although the design of GON involves integration, this operation has been efficiently converted into matrix computations, significantly reducing both inference and training times. In this case, the computational time for the finite element method is approximately 3 seconds. In comparison, GON achieves an inference time of 0.4 seconds, with a total training time of 23 minutes (1.4 seconds per epoch). For DeepONet, the inference time is 0.57 seconds, and the training time is 19 minutes. The FNO demonstrates an inference time of 0.22 seconds and a training time of 3 minutes, while the PINN requires 67.5 minutes for training. 
The computational speeds of GON is comparable to other deep learning methods and  outperforms the traditional FEM. Furthermore, GON inherently benefits from the superposition principle of Green’s function computation, making it naturally parallelizable. For larger-scale mesh problems, the block algorithm introduced in Section \ref{sec3} can be applied to further accelerate both the training and inference processes of GON.

\begin{table}[H]
    \centering
    \renewcommand{\arraystretch}{1.25}
    \caption{The hyper-parameters and performance of different baseline models on case of flat plane.}
    \begin{tabular}{c|c|c|l|c|c}
    \toprule
    Model & Epochs & Learning rate & Layers & Training error & Testing error\\
    \midrule 
    GON & 2000 & 0.001 & [6, 12, 12, 12, 1] & \textcolor{black}{$\mathbf{2.31\times10^{-4}}$} & \textcolor{black}{$\mathbf{5.22\times10^{-4}}$} \\
    \hline
    FNO & \textcolor{black}{500} & 0.001 & \textcolor{black}{2, 6, 6, 6, 6, 1} & \textcolor{black}{$1.33 \times 10^{-3}$} & \textcolor{black}{$2.89 \times 10^{-3}$} \\
    \hline
    \multirow{3}{*}{DeepONet} & \multirow{3}{*}{2000} & \multirow{3}{*}{0.001} & Trunk Net: [3, 12, 12, 12, 1] & \multirow{3}{*}{\textcolor{black}{$3.68\times10^{-2}$}} & \multirow{3}{*}{\textcolor{black}{$3.45\times10^{-2}$}} \\
        & &  &  Branch Net 1: [$N_{points}$, 12, 12, 12, 1] &\\
        & &  &  Branch Net 2: [$N_{points}$, 12, 12, 12, 1]   & \\
    \hline
    \multirow{3}{*}{PI-DeepONet} & \multirow{3}{*}{1000} & \multirow{3}{*}{0.001} & Trunk Net: [3, 12, 12, 12, 1]  & \multirow{3}{*}{\textcolor{black}{$4.02\times10^{-2}$}} & \multirow{3}{*}{\textcolor{black}{$3.93\times10^{-2}$}} \\
        & &  & Branch Net 1: [$N_{points}$, 12, 12, 12, 1] &   \\
        & &  &  Branch Net 2: [$N_{points}$, 12, 12, 12, 1]   & \\
    \hline
    PINN  & 15000 & 0.001 & [3, 12, 12, 12, 1] & - & \textcolor{black}{$6.54\times10^{-2}$} \\
    \bottomrule
    \end{tabular}
    \label{tab:diffusion}
\end{table}

\textbf{(b) Reaction-diffusion equations case on a pipe.} 
As shown in Figure \ref{fig:case_diffusion}, in this case, the diffusion coefficient is set as a heaviside function $a(\mathbf{x})=0.002(x < 0.08)+0.001(x \geq 0.08)$, and the force term \(f(\boldsymbol{x})\) is generated by the GRF with a wavelength parameter \(\lambda = 1\). Here, we set the diffusion coefficient as a step function to simulate a common mixing phenomenon between two substances in a micro pipe. The pipe has a radius of \textcolor{black}{0.08 m} and a length of \textcolor{black}{0.8 m}. The mesh of the pipe consists of 3,904 tetrahedral cells and 868 vertices within the domain, along with 740 triangular faces on the boundary. The dataset is acquired by \text{COMSOL Multiphysics®}. \textcolor{black}{The $L^2$ error relative to the reference mesh ($2.40 \times 10^{-3}\,\mathrm{m}$) is $1.56 \times 10^{-1}$.}

This case presents a higher level of complexity compared to the previous one due to the discontinuous diffusion coefficient. Notably, the performance of the FNO deteriorates in the presence of irregular geometries. For the complex modes in this scenario, while FNO, DeepONet, and PI-DeepONet successfully capture the main characteristics of the solution, they fail to accurately resolve fine-scale variations. The PINN struggles to converge, achieving an MSE of approximately 1.71 after 15,000 iterations. In contrast, the GON framework demonstrates superior performance by effectively capturing detailed variations, underscoring its robustness and versatility in solving reaction-diffusion equations on both regular and irregular computational domains with heterogeneous diffusion coefficients.

\begin{table}[H]
    \centering
    \renewcommand{\arraystretch}{1.25}
    \caption{The hyper-parameters and performance of different baseline models on case of flat plane.}
    \begin{tabular}{c|c|c|l|c|c}
    \toprule
    Model & Epochs & Learning rate & Layers & Training error & Testing error\\
    \midrule 
    GON & 7000 & 0.001 & [6, 24, 24, 24, 1] & \textcolor{black}{$\mathbf{4.43\times10^{-4}}$} & \textcolor{black}{$\mathbf{4.63\times10^{-4}}$} \\
    \hline
     FNO & \textcolor{black}{2000} & 0.001 & \textcolor{black}{[2, 12, 12, 12, 12, 1]} & \textcolor{black}{$1.72\times10^{-3}$} & \textcolor{black}{$2.14\times10^{-3}$} \\
    \hline
    \multirow{3}{*}{DeepONet} & \multirow{3}{*}{10000} & \multirow{3}{*}{0.001} & Trunk Net: [3, 24, 24, 24, 1] & \multirow{3}{*}{\textcolor{black}{$1.95\times10^{-3}$}} & \multirow{3}{*}{\textcolor{black}{$2.84\times10^{-2}$}} \\
        & &  &  Branch Net 1: [$N_{points}$, 24, 24, 24, 1] &\\
        & &  &  Branch Net 2: [$N_{points}$, 24, 24, 24, 1]   & \\
    \hline
    \multirow{3}{*}{PI-DeepONet} & \multirow{3}{*}{10000} & \multirow{3}{*}{0.001} & Trunk Net: [3, 24, 24, 24, 1]  & \multirow{3}{*}{\textcolor{black}{$4.72\times10^{-3}$}} & \multirow{3}{*}{\textcolor{black}{$4.63\times10^{-2}$}} \\
        & &  & Branch Net 1: [$N_{points}$, 24, 24, 24, 1] &   \\
        & &  &  Branch Net 2: [$N_{points}$, 24, 24, 24, 1]   & \\
    \hline
    PINN  & 15000 & 0.001 & [3, 24, 24, 24, 1] & - & $1.64$ \\
    \bottomrule
    \end{tabular}
    \label{tab:diffusion2}
\end{table}

\begin{figure}[H]
    \centering
    \includegraphics[width=0.85\linewidth]{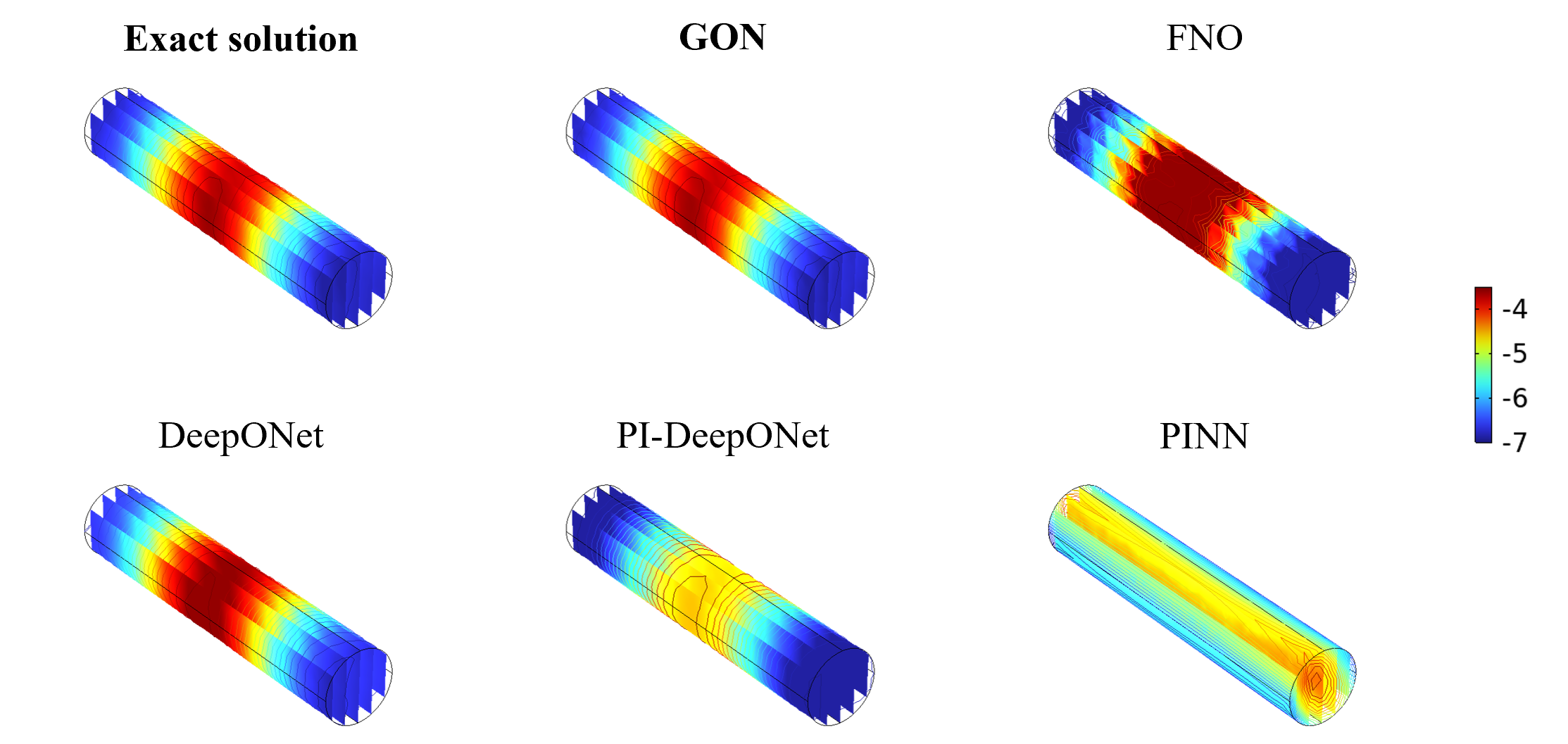}
    \caption{Case study for steady heterogeneous reaction-diffusion on flat plane: physical distribution and isotherm distribution from exact solution, and inference of GON, FNO, DeepONet, PI-DeepONet, PINN  (from left column to right column, from first row to second row).}
    \label{fig:diffusion2}
\end{figure}

\subsection{Case 3: Stokes equations}
We consider a classical benchmark problem in fluid dynamics: the 3D lid-driven cavity problem \cite{elman2014finite}. In this study, we aim to identify the matrix of Green's functions for Stokes flow \cite{boulle2022data, blake1971note}, modeled by the following system of equations over the domain $\Omega=[0,1]^3$:

\begin{equation}
    \left\{\begin{aligned}
    \mu \nabla^2 \mathbf{u}(\boldsymbol{x})-\nabla p(\boldsymbol{x})=\mathbf{f}(\boldsymbol{x}), \\
    \nabla \cdot \mathbf{u}(\boldsymbol{x})=0.
    \end{aligned}\right.
\end{equation}

\begin{figure}[H]
    \centering
    \includegraphics[width=0.9\linewidth]{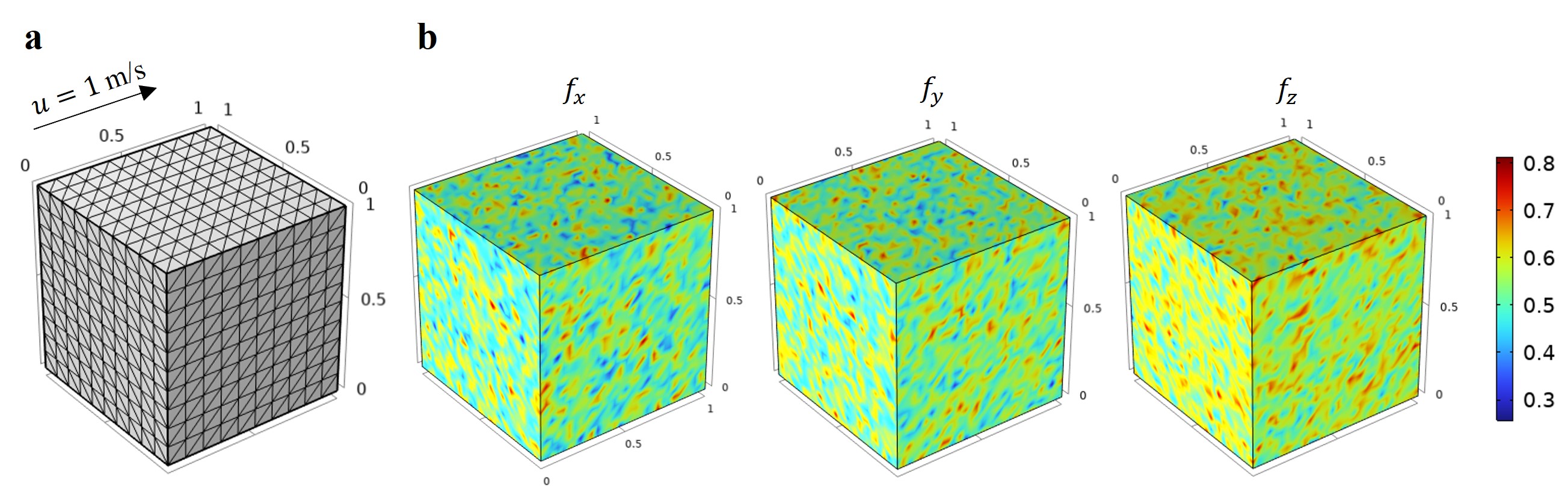}
    \caption{\textbf{a}. The simulation setting of the Stokes case; \textbf{b}. An example of the applied body force generated by GRF.}
    \label{fig:case_stokes}
\end{figure}

The governing equations describe a coupled system involving both velocity $\mathbf{u}=\left(u_x, u_y, u_z\right)$ and pressure fields $p$. $\mathbf{f}=\left(f_x, f_y, f_z\right)$ is an applied body force, and $\mu=1 / 100$ is the dynamic viscosity.
As shown in Figure \ref{fig:case_stokes}, the fluid velocity obeys no-slip boundary conditions on all walls except the top wall, where $\mathbf{u}=(1,0,0)$. The forcing term is generated using a GRF with a wavelength parameter $\lambda=0.1$. Figure \ref{fig:case_stokes} provides an illustration of the applied body force. The mesh is uniformly discretized into tetrahedral cells, containing 1331 vertices, 6000 tetrahedral cells within the domain, and 1200 triangular faces on the boundaries.
Based on this simulation setup, we generate training and testing datasets for all models by calculating the corresponding velocity solutions ($\mathbf{u}$) for various random body forces ($\mathbf{f}$) using \text{COMSOL Multiphysics®}. \textcolor{black}{The reference solution is computed on a mesh with an average size of $1.50 \times 10^{-2}\,\mathrm{m}$, with an $L^2$ norm error of approximately $3.71 \times 10^{-1}$ on this case.}

\begin{figure}[H]
    \centering
    \includegraphics[width=0.9\linewidth]{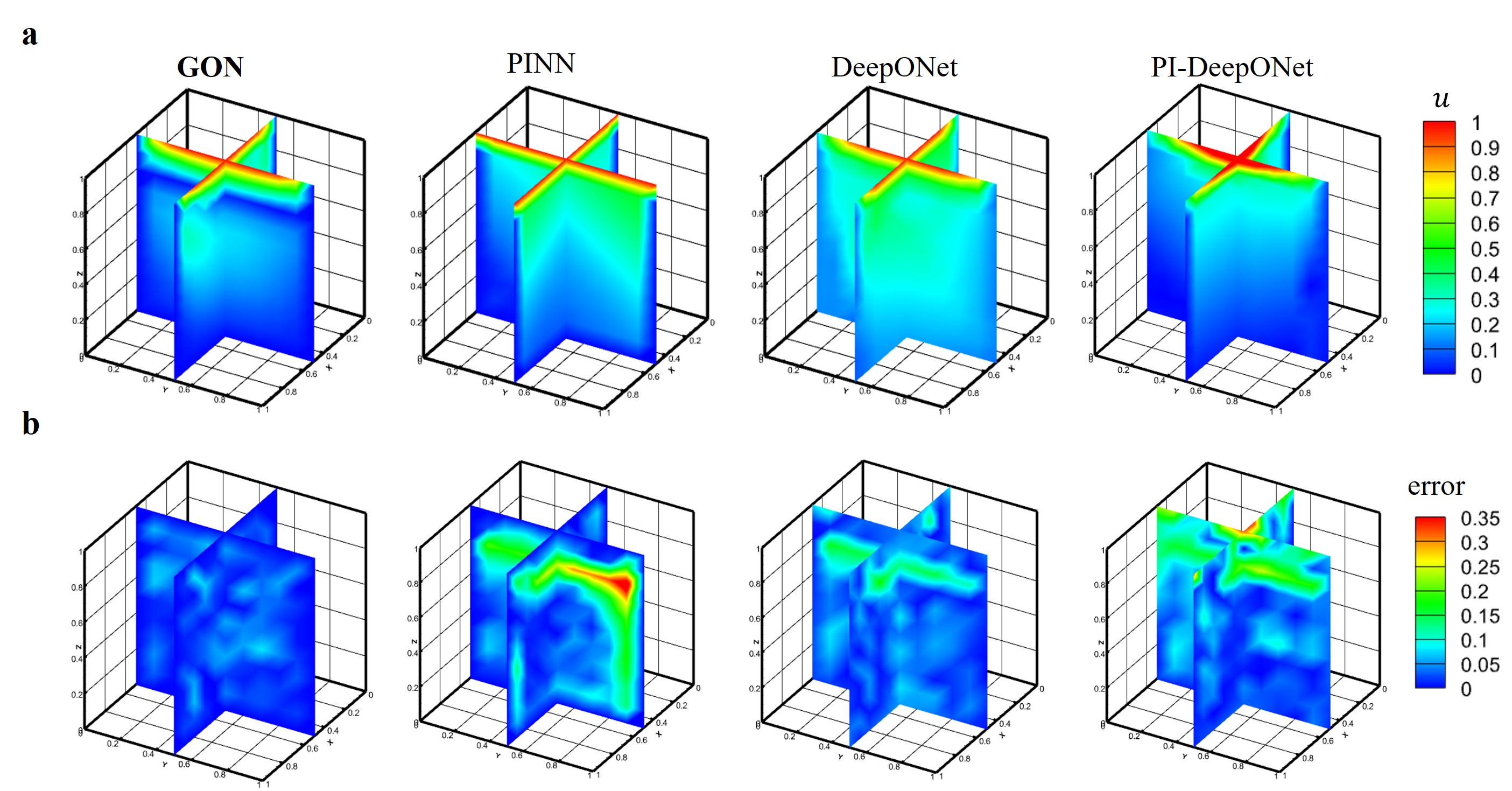}
    \caption{Case study for stokes cavity flows: \textbf{a} velocity magnitude from inference of GON, PINN, DeepONet, PI-DeepONet, FNO (from left to right); \textbf{b} Point-wise error of velocity magnitude ($|\hat{u}-u|$) by GON, PINN, DeepONet, PI-DeepONet, FNO (from left to right); \textbf{c} Stream traces calculated by results of GON, PINN, DeepONet, PI-DeepONet, FNO (from left to right).}
    \label{fig:stokes_res}
\end{figure}

Table \ref{tab:stokes} and Figure \ref{fig:stokes_res} provide detailed comparisons between GON and the baseline models.
GON achieved the lowest \(L_2\) error of $\mathbf{5.83\times10^{-4}}$ on the testing set. Specifically, GON's input channel consists of 6 units, accepting $(\mathbf{x}, \mathbf{\xi})$ as input. For this Stokes problem, a $3\times3$ network matrix was trained within GON to learn the correlations between $(u_x, u_y, u_z)$ and $(f_x, f_y, f_z)$, as outlined in Section \ref{sec2}. As shown in Figure \ref{fig:stokes_res} \textbf{a} and \textbf{b}, GON demonstrated superior performance over PINN, DeepONet, PI-DeepONet, and FNO in calculating the velocity field for new force terms. Furthermore, Figure \ref{fig:stokes_res} \textbf{c} reveals that GON successfully captures the primary flow characteristics of the lid-driven cavity flow, notably the large central vortex. In contrast, despite their relatively low errors, PINN, DeepONet, and PI-DeepONet fail to fully capture the vortex patterns within the cavity as effectively as GON.

In this work, we apply the GON to realize rapid solution of specific PDEs under varying boundary conditions and source terms. Our approach can be further extended to solve the Navier-Stokes equations by leveraging GON for the efficient solution of the Poisson equation, thereby accelerating the iterative process of solving the Navier-Stokes system. Furthermore, GON holds significant potential for applications in scenarios requiring multiple evaluations of the forward problem, such as uncertainty quantification, inverse problems, and optimization tasks.

\begin{table}[H]
    \centering
    \caption{The training and model hyper-parameters of different baseline models.}
    \begin{tabular}{c|c|c|l|c|c}
    \toprule
    Model & Epochs & Learning rate & Layers & Training error & Testing error\\
    \midrule 
    GON & 2000 & 0.001 & [6, 12, 24, 12, 1] & $mathbf{5.71\times10^{-4}}$ & $\mathbf{5.83\times10^{-4}}$ \\
    FNO & \textcolor{black}{500} & 0.001 & \textcolor{black}{[2, 12, 12, 12, 12, 3]} & \textcolor{black}{$1.36\times10^{-3}$} & \textcolor{black}{$2.21\times10^{-3}$} \\
    \multirow{3}{*}{DeepONet} & \multirow{3}{*}{2000} & \multirow{3}{*}{0.001} & Trunk Net: [3, 12, 24, 12, 4] & \multirow{3}{*}{$2.15\times10^{-3}$} & \multirow{3}{*}{$2.17\times10^{-3}$} \\
        & &  &  Branch Net 1: [3$N_{points}$, 12, 24, 12, 4] &\\
        & &  &  Branch Net 2: [$N_{points}$, 12, 24, 12, 4]   & \\
    \multirow{3}{*}{PI-DeepONet} & \multirow{3}{*}{2000} & \multirow{3}{*}{0.001} & Trunk Net: [3, 12, 24, 12, 4]  & \multirow{3}{*}{$5.59\times10^{-3}$} & \multirow{3}{*}{$5.31\times10^{-3}$} \\
        & &  & Branch Net 1: [3$N_{points}$, 12, 24, 12, 4] &   \\
        & &  &  Branch Net 2: [$N_{points}$, 12, 24, 12, 4]   & \\
    PINN  & 2000 & 0.001 & [3, 12, 24, 12, 4] & - & $1.08\times10^{-2}$ \\
    \bottomrule
    \end{tabular}
    \label{tab:stokes}
\end{table}

\section{Conclusions\label{sec5}}
In this work, we propose GON, a novel framework inspired by Green's functions, designed to address key limitations of existing methods like PINN and DeepONet. Unlike PINN, GON can directly compute new solutions for varying boundary conditions and source terms without retraining. Compared to DeepONet, GON offer superior interpretability and enhanced approximation capabilities. 
To evaluate its performance, we conducted experiments on three classical equations: the heat equation, the reaction-diffusion equation, and the Stokes equation. These tests included scenarios with varying boundary conditions, source terms, and both homogeneous and heterogeneous setups. GON consistently outperformed state-of-the-art methods.
The GON framework provides flexibility in handling user-defined meshes, boundary conditions, and initial conditions, making it highly accessible to engineers accustomed to traditional computational engineering simulations. Looking ahead, we plan to extend GON to tackle nonlinear, multiphysics coupled equations.
\textcolor{black}{A potential approach, as proposed in \cite{gin2021deepgreen}, involves leveraging neural networks to map nonlinear equations into a linear space, solve them in the transformed domain, and revert the solution to the original space. We are actively investigating this and other methodologies to systematically extend our framework to nonlinear problems.}

\section*{\textcolor{black}{Acknowledgments}}
This work    was    supported    by Natural      Science     Foundation     of    Ningbo     of     China     (No. 2023J027),  China Meteorological    Administration     under     Grant     QBZ202316, the    National Natural Science Foundation of    China (Grant    No.62106116), as well as by the High Performance Computing Centers at Eastern Institute of Technology, Ningbo, and Ningbo Institute of  Digital Twin.

\bibliographystyle{unsrt}
\bibliography{aipsamp}

\newpage
\appendix
\renewcommand{\thefigure}{A\arabic{figure}}
\setcounter{figure}{0}
\renewcommand{\thetable}{A\arabic{table}}
\setcounter{table}{0}

{\color{black}
\section*{Appendix}
\subsection*{A.1 Effect of quadrature rule}
To investigate the impact of the number of integration points on the accuracy of the learned Green's function, we conducted additional numerical experiments on \textit{Case 1: Steady Heat Conduction}. 
To more clearly observe the influence of the quadrature rule, the gradient of the Green’s function with respect to the source location is computed via automatic differentiation rather than learned through the Branch Net. This choice eliminates the potential approximation error from learning the gradient, ensuring that any observed differences in accuracy stem primarily from the number of integration points used.
We evaluated the relative \( L_2 \) errors using Eq. \eqref{eq10}, with varying numbers of Gaussian quadrature points. The results are summarized in Table A1. 

For completeness, the corresponding Gaussian quadrature formulas and their associated weights for different numbers of integration points are provided below:

(1) \textbf{Quadrature rule 1} (1-point quadrature rule on $E_m$, 1-point quadrature rule on $T_l$):
\[
\begin{aligned}
&\mathbf{\eta} = [\eta_{i1}]
= \begin{bmatrix}
\frac{1}{3}  \\[6pt]
\frac{1}{3} \\[6pt]
\frac{1}{3} 
\end{bmatrix}, \quad
\mathbf{w} = \begin{bmatrix}
1
\end{bmatrix}.\\
&\mathbf{\zeta} = [\zeta_{i1}]
= \begin{bmatrix}
\frac{1}{4} \\[6pt]
\frac{1}{4} \\[6pt]
\frac{1}{4} \\[6pt]
\frac{1}{4}
\end{bmatrix}, \quad
\mathbf{w} = \begin{bmatrix}
1
\end{bmatrix}.
\end{aligned}
\]

(2) \textbf{Quadrature rule 2} (3-point quadrature rule on $E_m$, 4-point quadrature rule on $T_l$):
\[
\begin{aligned}
&\mathbf{\eta} = [\eta_{i1},\eta_{i2},\eta_{i3}]
= \begin{bmatrix}
\frac{2}{3} & \frac{1}{6} & \frac{1}{6} \\[6pt]
\frac{1}{6} & \frac{2}{3} & \frac{1}{6} \\[6pt]
\frac{1}{6} & \frac{1}{6} & \frac{2}{3} 
\end{bmatrix}, \quad
\mathbf{w} = \begin{bmatrix}
\frac{1}{3} \\[6pt]
\frac{1}{3} \\[6pt]
\frac{1}{3} 
\end{bmatrix}.\\
&\mathbf{\zeta} = [\zeta_{i1},\zeta_{i2},\zeta_{i3}]
= \begin{bmatrix}
0.5854 & 0.1382 & 0.1382 \\[6pt]
0.1382 & 0.5854 & 0.1382 \\[6pt]
0.1382 & 0.1382 & 0.5854 \\[6pt]
0.1382 & 0.1382 & 0.1382
\end{bmatrix}, \quad
\mathbf{w} = \begin{bmatrix}
\frac{1}{4} \\[6pt]
\frac{1}{4} \\[6pt]
\frac{1}{4} \\[6pt]
\frac{1}{4}
\end{bmatrix}.
\end{aligned}
\]

(3) \textbf{Quadrature rule 3} (6-point quadrature rule on $E_m$, 5-point quadrature rule on $T_l$):
\[
\begin{aligned}
&\mathbf{\eta} = [\eta_{i1},\eta_{i2},\eta_{i3},\eta_{i4},\eta_{i5},\eta_{i6}]
= \begin{bmatrix}
0.8168 & 0.0916 & 0.0916 & 0.1081 & 0.4459 & 0.4459 \\[6pt]
0.0916 & 0.8168 & 0.0916 & 0.4459 & 0.1081 & 0.4459 \\[6pt]
0.0916 & 0.0916 & 0.8168 & 0.4459 & 0.4459 & 0.1081
\end{bmatrix}, \quad
\mathbf{w} = \begin{bmatrix}
0.1099 \\[6pt]
0.1099 \\[6pt]
0.1099 \\[6pt]
0.2234 \\[6pt]
0.2234
\end{bmatrix}.\\
&\mathbf{\zeta} = [\zeta_{i1},\zeta_{i2},\zeta_{i3}, \zeta_{i4}, \zeta_{i5}]
= \begin{bmatrix}
\frac{1}{4} & \frac{1}{2} & \frac{1}{6} & \frac{1}{6} & \frac{1}{6} \\[6pt]
\frac{1}{4} & \frac{1}{6} & \frac{1}{2} & \frac{1}{6} & \frac{1}{6} \\[6pt]
\frac{1}{4} & \frac{1}{6} & \frac{1}{6} & \frac{1}{2} & \frac{1}{6} \\[6pt]
\frac{1}{4} & \frac{1}{6} & \frac{1}{6} & \frac{1}{6} & \frac{1}{2}
\end{bmatrix}, \quad
\mathbf{w} = \begin{bmatrix}
-\frac{4}{5} \\[6pt]
\frac{9}{20} \\[6pt]
\frac{9}{20} \\[6pt]
\frac{9}{20} \\[6pt]
\frac{9}{20}
\end{bmatrix}.
\end{aligned}
\]

As demonstrated in Table A1, it is evident that increasing the number of integration points enhances accuracy, as seen from the decreasing training and testing errors. Specifically, Quadrature rule 1 (1-point quadrature rule on \( E_m \), 1-point quadrature rule on \( T_l \)) exhibits the highest errors, while Quadrature rule 3 (6-point quadrature rule on \( E_m \), 5-point quadrature rule on \( T_l \)) achieves the lowest. However, this improvement comes at the cost of increased GPU memory usage. Quadrature rule 2 (3-point quadrature rule on \( E_m \), 4-point quadrature rule on \( T_l \)) provides a good trade-off, balancing accuracy and efficiency, making it the preferred choice for our integration scheme.
}

\begin{table}[H]
    \centering
    {\color{black}
    \caption{MSE error calculated by Eq. \eqref{eq10} for Case 1 with respect to the different number of Gaussian quadrature points used.}
    \begin{tabular}{c|cccc}
        \hline Quadrature rule & Training error & Testing error & Memory usage & Inference time  \\
        \hline 
        Quadrature rule 1 & $8.20\times10^{-5}$ & $1.12\times10^{-4}$ & 839 MiB & 0.098 s \\
        Quadrature rule 2 & $7.56\times10^{-5}$ & $9.74\times10^{-5}$ & 963 MiB & 0.11 s \\
        Quadrature rule 3 & $6.49\times10^{-5}$ & $8.88\times10^{-5}$ & 1461 MiB & 0.12 s \\
        \hline
    \end{tabular}
    }
    \label{tab:quadrature}
\end{table}

{\color{black}
\subsection*{A.2 Convergence behavior: BsNN vs. FNN}
To further examine the convergence characteristics of the BsNN, we conducted additional experiments on \textit{Case 1: Steady Heat Conduction}, comparing its performance with that of a standard FNN. Two configurations were examined: (a) where the gradient of the Green’s function is approximated by the Branch Net, and (b) where the gradient is obtained directly via automatic differentiation.

In configuration (a), as shown in Figure A1(a), both BsNN and FNN exhibit convergence. However, BsNN demonstrates significantly faster convergence, reaching a relative \(L_2\) error of \(4.55 \times 10^{-5}\) within 3,000 training iterations, whereas FNN plateaus at a higher error of \(7.77 \times 10^{-3}\). This improvement might be attributed to BsNN’s superior capability in capturing the diagonal singularity structure inherent to Green’s functions.

In configuration (b), where gradient information is derived solely from automatic differentiation rather than learned by the Branch Net, the robustness of BsNN becomes more pronounced. As shown in Figure A1(b), FNN fails to converge and exhibits an increasing loss trend throughout training, ultimately resulting in a relative \(L_2\) error of \(1.54 \times 10^{-2}\). In contrast, BsNN maintains stable convergence and achieves a final error of \(8.75 \times 10^{-5}\).

These results demonstrate that BsNN not only improves convergence speed relative to FNN, but also enhances training stability under more challenging conditions, underscoring its effectiveness in modeling Green’s functions with singular behaviors.
}

\begin{figure}[H]
    \centering
    \includegraphics[width=1\linewidth]{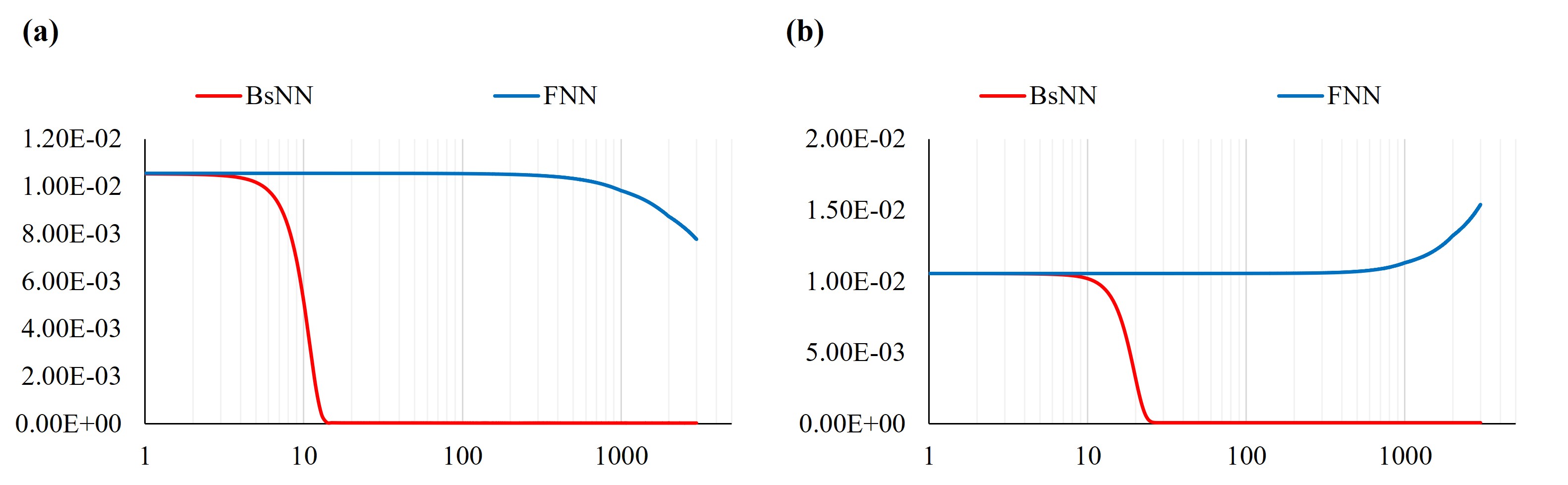}
    \caption{Convergence behavior of BsNN and FNN on \textit{Case 1} \textcolor{black}{(x-axis in log scale)}: (a) Performance comparisons when the gradient of the Green’s function is approximated by the Branch Net; (b) Performance comparisons when the gradient is obtained directly via automatic differentiation.}
    \label{fig:bsnn_fnn}
\end{figure}

{\color{black}
\subsection*{A.3 Effect of Branch Net}
To investigate the impact of the Branch Net on training dynamics, we compare the convergence behavior of two model variants: one with the Branch Net and one without. Figure A2 illustrates the training loss curves for both configurations on two representative cases: Case 3 with a regular computational domain (panel (a)) and Case 1 with an irregular computational domain (panel (b)).

In both scenarios, the integration of the Branch Net evidently improves the convergence speed and reduces the final training error. Specifically, in Case 3, the final loss is reduced from \(7.78 \times 10^{-4}\) to \(6.31 \times 10^{-4}\) when the Branch Net is used. The improvement is even more pronounced in Case 1, where the presence of geometric irregularities poses greater challenges for learning. In this case, the final loss is nearly halved, from \(8.83 \times 10^{-5}\) to \(4.60 \times 10^{-5}\).  \textcolor{black}{While the convergence curve in Case 1 exhibits a more rapid decline, this behavior likely stems from differences in the underlying PDEs.}

These results demonstrate that the Branch Net enhances the model’s expressive capacity and facilitates more efficient learning, particularly in domains with complex geometries. This design choice proves beneficial for both convergence speed and solution accuracy.
}

\begin{figure}[H]
    \centering
    \includegraphics[width=1\linewidth]{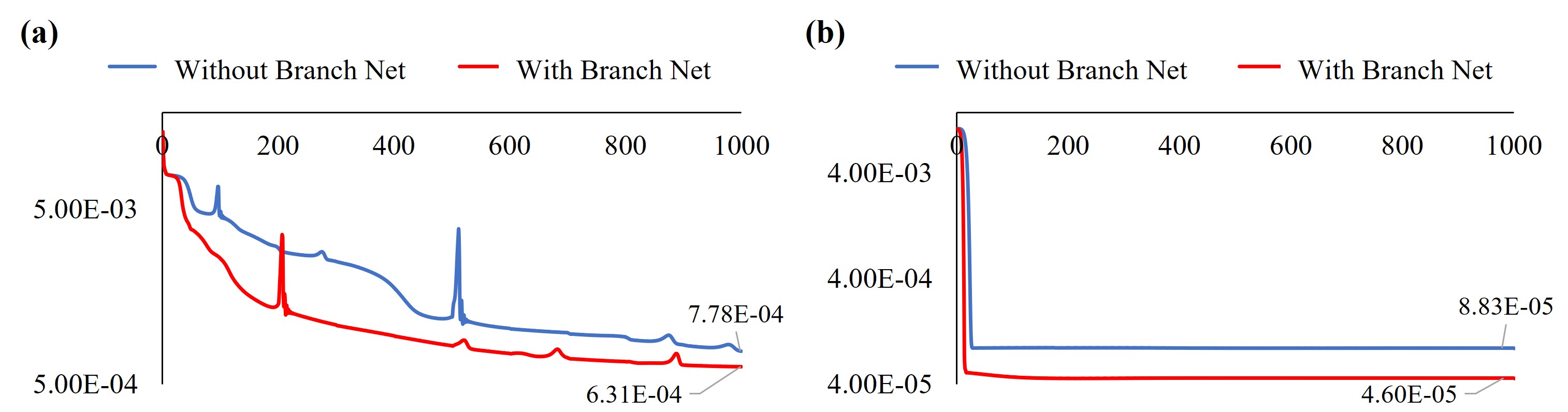}
    \caption{\textcolor{black}{Convergence behavior with and without the approximation by a Branch Net: (a) Comparison of training loss on Case 3 with a regular computational domain; (b) Comparison of training loss on Case 1 with an irregular computational domain.}}
    \label{fig:gradient}
\end{figure}

{\color{black}
\subsection*{A.4 $H^1$ semi-norm error for Case 1-Case 3}
To provide a more stringent assessment of approximation quality, we compute the \( H^1 \) semi-norm errors, which focused on gradient discrepancies, for all three benchmark cases (Case 1–3), as the \( H^1 \) norm is particularly relevant for PDE problems involving diffusion or conduction. Specifically, we evaluate the semi-norm  
    $$
    |u-\hat{u}|_{H^1}=\left(\int_{\Omega}|\nabla (u(x)-\hat{u})|^2 d x\right)^{1 / 2}
    $$
where \( u \) denotes the ground-truth solution and \( \hat{u} \) the model-predicted solution.

As summarized in following Tables, the trends observed in the \( H^1 \) semi-norm errors are consistent with those reported using the \( L^2 \) norm. These results further confirm the superior accuracy of our proposed method across different problem settings. To avoid interrupting the flow of the main manuscript, we present these detailed error metrics here in the appendix.
}

    \begin{table}[H]
    \centering
    {\color{black}
    \caption{Case 1. Model performance comparison evaluated by $H^1$ semi norm error.}
    \label{tab:case1}
    \begin{tabular}{lcc}
    \toprule
    \textbf{Model} & \textbf{Training set} & \textbf{Testing set} \\
    \midrule
        GON        & 0.0129 & 0.0149 \\
        FNO        & 0.0124 & 0.0185 \\
        DeepONet   & 0.0215 & 0.0295 \\
        PIDeepONet & 0.0295 & 0.0308 \\
        PINN       & {--}   & 0.0526 \\ 
    \bottomrule
    \end{tabular}
    }
    \end{table}
    \begin{table}[H]
    \centering
    {\color{black}
    \caption{Case 2 (a). Model Performance Comparison evaluated by $H^1$ semi norm error.}
    \label{tab:case2a}
    \begin{tabular}{lcc}
    \toprule
    \textbf{Model} & \textbf{Training set} & \textbf{Testing set} \\
    \midrule
        GON        & 0.1896 & 0.1836 \\
        FNO        & 0.1855 & 0.1864 \\
        DeepONet   & 0.8590 & 0.7297 \\
        PIDeepONet & 0.9084 & 0.7548 \\
        PINN       & {--}   & 0.3175 \\
    \bottomrule
    \end{tabular}
    }
    \end{table}
    \begin{table}[H]
    \centering
    {\color{black}
    \caption{Case 2 (b). Model Performance Comparison evaluated by $H^1$ semi norm error.}
    \label{tab:case2b}
    \begin{tabular}{lcc}
    \toprule
    \textbf{Model} & \textbf{Training set} & \textbf{Testing set} \\
    \midrule
        GON        & 0.0988  & 0.0999  \\
        FNO        & 0.3070  & 0.6638  \\
        DeepONet   & 0.7826  & 0.7089  \\
        PIDeepONet & 1.3529  & 0.7818  \\
        PINN       & {--}    & 2.6713  \\
    \bottomrule
    \end{tabular}
    }
    \end{table}
    \begin{table}[H]
    \centering
    {\color{black}
    \caption{Case 3. Model Performance Comparison evaluated by $H^1$ semi norm error.}
    \label{tab:case3}
    \begin{tabular}{lcc}
    \toprule
    \textbf{Model} & \textbf{Training set} & \textbf{Testing set} \\
    \midrule
        GON        & 0.0429 & 0.0447 \\
        FNO        & 0.0479 & 0.0779 \\
        DeepONet   & 0.0542 & 0.0986 \\
        PIDeepONet & 0.2638 & 0.1966 \\
        PINN       & {--}   & 0.5050 \\
    \bottomrule
    \end{tabular}
    }
    \end{table}

\end{document}